\documentclass[pre,preprint]{revtex4}

\usepackage{latexsym,dsfont,array,layout,mathrsfs}
\usepackage{amsfonts}
\usepackage{amsbsy}
\usepackage{amsmath}
\usepackage{amssymb}
\usepackage{theorem}
\usepackage{graphics}
\usepackage{epsfig}
\usepackage{graphicx}
\usepackage{bm}
\usepackage{latexsym}

\newcommand{\one}{\openone}

\newcommand{\ket}[1]{| #1 \rangle}  
\newcommand{\bra}[1]{\langle  #1 |} 
\newcommand{\natuerl}{{\mathbb N}}
\newcommand{\Tr}{\mbox{Tr}}

\newcommand{\Ud}{\hat{U}^\dagger}
\newcommand{\vac}{\ket{\mbox{vac}}}
\theoremstyle{break}

\begin{document} 
\title{Quantum states prepared by 
realistic entanglement swapping} 
\date{March 27, 2009}
\author{Artur Scherer\footnote{the corresponding author:
ascherer@ucalgary.ca}}
\author{Gina Howard}
\author{Barry C.\ Sanders}
\author{Wolfgang Tittel}
\affiliation{Institute for Quantum Information Science, University of
Calgary, Alberta T2N 1N4, Canada}
       
\begin{abstract}

Entanglement swapping between photon pairs is a fundamental building block
in schemes using quantum relays or quantum repeaters to overcome the range
limits of long distance quantum key distribution. We develop a closed-form 
solution for the actual quantum states prepared by realistic entanglement 
swapping, which takes into account experimental deficiencies due to
inefficient detectors, detector dark counts and multi-photon-pair 
contributions of parametric down conversion sources. 
We investigate how the entanglement present in the final state 
of the remaining modes is affected 
by the real-world imperfections. To test the predictions of our theory,
comparison with previously published experimental entanglement swapping is provided.
\vspace*{2mm}

\noindent
PACS numbers: 03.67.-a, 03.67.Bg, 03.67.Dd, 03.67.Hk, 42.50.Ex    \\
\end{abstract}

\maketitle

\vspace{4mm}

\noindent
\section{Introduction}

Quantum cryptographic communication and quantum key distribution technologies
have matured to a level sufficient for commercial applications. Yet, distance limits 
impact on their usefulness. To date, the only realistic proposals for long distance quantum 
cryptography are still based on optical systems. Light is an optimal 
candidate to be a carrier of quantum information because photonic quantum 
states are durable due to their generally weak interaction with the environment 
and are conveniently manipulated by means of linear optics and
photon-detection. Moreover, photons are the fastest and one of the simplest 
physical systems for encoding quantum information. The technological challenge 
is to establish transmission channels over long distances with a high 
signal-to-noise ratio using real-world optical fiber settings 
or free space. Long distance quantum communication (LDQC) is hampered by a 
significant loss of photons with distance traveled. 
In particular, loss of photons due to absorption during transmission in fibers 
is characterized by the exponential rule $t=10^{-\alpha d/10}\;,$
where $t$ is the transmission coefficient, $d$ is the distance
traveled and $\alpha$ is the loss coefficient of the transmission
medium in units of dB. A further limitation 
to the development of quantum  communication over long distances 
is a constant detector noise level dominating over the exponential decrease 
of the signal. In an effort to overcome these obstacles and range  
limits of LDQC, quantum repeaters~\cite{BriegelDuerCiracZoller1998,Duan2001} 
or quantum relays~\cite{Gisin2002,Waks2002,Jacobs2002,Riedmatten2004,Collins2005} 
have been proposed, which comprise entanglement swapping~\cite{Zukowski1993} 
as a fundamental building block. 

In principle, quantum repeaters enable any distance
to be achieved. The basic idea of a quantum repeater is to split 
the long distance quantum channel into shorter segments and to 
distribute entanglement between the end nodes of these segments.  
Then, after purifying the noisy entanglement for each segment, 
the entanglement is extended over adjacent segments by means 
of entanglement swapping. The purification procedure is repeated for the extended
segments, and the whole protocol reiterated until
high-purity entanglement is established between the end
points of the channel. A quantum relay works in a similar way 
as the quantum repeater, but without the entanglement purification procedure 
and without quantum memories. This makes it much more feasible 
as compared to the repeater, but does not allow achieving arbitrary 
distances~\cite{Collins2005}. With both schemes    
the signal-to-noise ratio can be appreciably increased.
However, experimental realization~\cite{Collins2005} 
suffers from a number of imperfections, including imperfect sources
of entangled pairs and imperfect detectors. 

The impact of experimental deficiencies on the security of
a quantum channel as well as on sifted and secret key rates
in quantum key distribution (QKD) is of considerable relevance
and has been the objective of a number of recent investigations. 
In~\cite{Brassard2000}
Brassard et al.\ showed that channel losses, a realistic detection
process comprising detector inefficiencies and dark counts, and
imperfections in the qubit source drastically impair the feasibility
of QKD over long distances. In particular, the implications of
using attenuated laser pulses instead of idealized single-photon
on-demand sources were examined, and it was shown that unconditional security
is very difficult to achieve in long distance QKD based on a
BB84 protocol~\cite{BB84} with such weak laser pulses. In the 
same work Brassard and coworkers obtained a more optimistic 
performance for QKD schemes based on parametric down-conversion (PDC)
sources~\cite{Ekert92,Sergienko99}. The consequences of using 
probabilistic photon-pair sources (as realized by PDC) instead 
of (non-existing) single-pair on-demand sources for quantum 
communication including entanglement 
based QKD have recently further been investigated, 
see \cite{Ma2007,Marcikic2002,Riedmatten2003}. 

For LDQC employing quantum repeaters or relays it is particularly 
important to examine the issue of how the entangled quantum states 
after an entanglement swapping operation are affected by experimental 
imperfections. It is clear that due to these imperfections the 
actual quantum states deviate from desired Bell states and have 
to be described by some mixed states. The impact of transmission losses 
and detector inefficiencies as well as dark counts on the performance of 
quantum relays has been recently examined by Collins et.\ al.\ in~\cite{Collins2005}. 
However, the probabilistic nature of photon-pair sources has not been
considered in their work. As will be explained in the next section, 
the probabilistic nature of PDC also involves the 
possibility of multi-pair generation. 
Depending on the \lq\lq brightness'' of the sources, the emission of two 
(or even more) independent pairs of entangled photons 
from the same PDC source at a time  becomes a more or less 
significant event leading to faulty detection clicks
and incorrect conclusions with regard to entanglement and 
correlations. Thus, the multi-pair nature of PDC sources 
impairs a high fidelity realization of entanglement swapping. 
The investigation of the issue as to what extent the inevitable multi-pair
contributions of PDC sources impinge on the performance of quantum communication 
protocols based on entanglement swapping, is the main motivation for the
research presented in this article. 
The effect of multi-excitation events in PDC in addition to 
detector imperfections and transmission losses on quantum
repeater performance has been examined perturbatively 
in the context of single-rail entanglement by 
Brask and S{\o}rensen in \cite{BraskSorensen2008}. 
Furthermore, the atom-light entangled states produced by Stokes scattering in
the DLCZ-scheme~\cite{Duan2001} are very similar to the light-light
entangled states produced in a non-degenerate PDC process, as both are 
two-mode squeezed states. In works by Jiang 
{\em et al}.\ ~\cite{JiangTaylorLukin2007} and Zhao {\em et al}.\
~\cite{Zhao-et-al2007}, 
which extend the original DLCZ scheme to dual-rail
entanglement, the consequence of multi-excitation events has also been treated 
in a perturbative way. In our work we choose a different approach, which is 
{\em non-perturbative} and essentially simpler, as it uses only the very basic 
toolbox of quantum theory and the principle of Bayesian inference. 

In this article, we elaborate on how the entangled quantum states after 
entanglement swapping are affected by experimental imperfections, 
particularly the multi-pair contributions of PDC sources.
We provide a non-perturbative theory for realistic entanglement swapping   
with imperfect photon-pair sources as well as imperfect detectors. 
In particular, we incorporate the multi-pair nature of PDC, 
transmission losses, and detector inefficiencies and dark counts 
  non-perturbatively. 
Our theory enables us to obtain a closed-form analytic solution for 
the resultant mixed entangled quantum states after a real-world       
(noisy) entanglement swapping operation. To test our theory, we compare 
its predictions with actual experiments on entanglement swapping, 
that have been previously published elsewhere.
For this purpose, we derive a closed-form expression for the 
probability of four-fold coincidences of four detectors, 
two for the Bell-state measurement and two for monitoring 
the remaining two entangled modes, one on each side, 
depending on variable polarization directions of analyzers. 
This result allows us to calculate numerically the visibility 
of four-fold coincidences for arbitrary parameter values 
characterizing real-world imperfections of the PDC sources 
and detectors. Finally, we inspect how the entanglement present 
in the final state of the remaining modes is 
affected by the practical deficiencies. 
The analysis makes it possible to suggest
the implications of the imperfections on schemes using entanglement
swapping as a fundamental tool, 
and to optimize parameter settings for maximum performance.

In addition to the imperfections of sources and detectors, 
further problems are encountered in a realistic entanglement 
swapping process. Imperfections of temporal overlap of the 
light fields on a beamsplitter as well as spectral mode mismatch 
constitute inevitable practical difficulties deteriorating the 
performance. Moreover, the entangled photon pairs prepared by two 
sources are expected not to be of the same quality. 
All of these problems would complicate the 
analysis and are not taken into account in the present study, 
but are planned to be included in our future work. 
Here, we would like to focus on implications of imperfect 
sources and imperfect detectors only. In this way, our considerations 
provide a very useful upper bound on the amount of entanglement 
after swapping. As for the influence of mode mismatch on the 
performance, we would like to refer the reader to models 
developed in \cite{Oezdemir2002,Rohde2006,Jia2005}. 

Our work extends the previous challenge~\cite{Brassard2000} 
to the security of QKD protocols using imperfect sources 
as well as the investigations in 
\cite{Collins2005,BraskSorensen2008,JiangTaylorLukin2007,Zhao-et-al2007}.
Efforts toward establishing reliable
transmission of quantum systems over arbitrary distances will
benefit from our careful analysis presented here.

This article is organized as follows. In section~\ref{OurTheory} 
we develop our theory of real-world entanglement swapping. 
A mathematical description of imperfect photon-pair sources 
and imperfect detectors is provided. Using these models we 
derive a closed-form solution for the quantum states 
after a realistic entanglement swapping operation. 
In section~\ref{ComparisonWithExperimentalSwapping} we 
apply our theory to making predictions with regard to 
entanglement verification in terms of the visibility of 
four-fold coincidences and compare 
our numerical results with experimental entanglement 
swapping. We proceed with a discussion of the impact of experimental 
deficiencies on the entanglement of the resultant 
quantum state.  We conclude with a brief summary and 
suggestions for future research  in section~\ref{Conclusion}.

\section{A theory for practical entanglement swapping}
\label{OurTheory}

\subsection{Physical situation and setting}

\begin{figure}[bt]
\center{\includegraphics[width=0.67\linewidth]{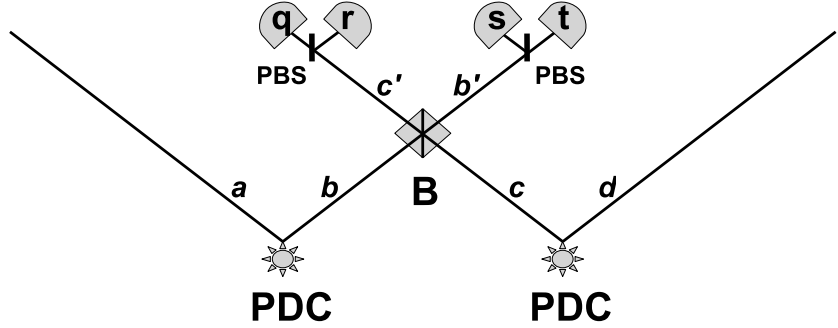}}
{\caption{Entanglement swapping of photon-polarization qubits,  
based on two imperfect parametric down
conversion sources (PDC) and a Bell-measurement with four imperfect
photon detectors. Four spatial modes are involved, labeled by $a$, $b$, $c$ 
and $d$. Two modes, one from the first and one from the second source, 
$b$ and $c$, respectively, are 
combined at a balanced beam-splitter (B). The exits of the latter, 
denoted by $b'$ and $c'$, respectively,   
are directed to polarizing beamsplitters (PBS) and then detected 
at four detectors: one for the $H$ and one for the $V$ polarization 
of each of the $c'$ and $b'$ modes. 
This set-up forms an interferometric
Bell-state measurement \cite{Weinfurter1994,Weinfurter1995}. 
The four detectors are inefficient photon detectors subject to  
dark counts. Their readout is denoted by $(qrst)$. Given this readout 
we are interested in the entangled quantum state of the remaining 
$a$ and $d$ modes depending on experimental parameters characterizing 
the deficiencies of the experiment. 
\label{Fig:entswap2}}}
\end{figure}

In this paper we develop our theory for entanglement swapping of 
photon-polarization qubits as an illustrative example. However, the main 
issues of the theory apply to any other photonic realization of qubits
\cite{TittelWeihs2001}    
and the results and implications are independent of the latter. 
The basic experimental situation is illustrated in Fig.~\ref{Fig:entswap2}. 
Two parametric down conversion sources emit photon pairs into spatial modes  
$a$, $b$, $c$ and $d$, where $a$ and $b$ correspond to the first and 
$c$ and $d$ to the second PDC source. In the ideal-case 
scenario, one entangled photon-pair is emitted into the 
$a$ and $b$ modes and another one into the $c$ and $d$ 
modes. For entanglement swapping, a joined Bell-state 
measurement is performed on the  $b$ and $c$ modes. 
This results in projecting the remaining  
modes $a$ and $d$ onto an entangled state, depending on 
the measurement readout of the Bell-state 
measurement. As a consequence, the photons in the 
outgoing $a$ and $d$ modes emerge entangled despite 
never having had an interaction with one another \cite{Zukowski1993}. 
The entanglement previously contained in the 
$a$ and $b$ and the $c$ and $d$ photon pairs is 
swapped to the $a$ and $d$ photon pair. 

As explained in Fig.~\ref{Fig:entswap2}, in the case of polarization qubits  
a Bell-state measurement 
consists in combining the $b$ and $c$ modes at a balanced beam-splitter,  
then directing its output modes $b'$ and $c'$ 
to polarizing beamsplitters (PBS) and 
finally detecting the four alternatives $c'_{\mbox{\tiny{H}}}$,
$c'_{\mbox{\tiny{V}}}$, $b'_{\mbox{\tiny{V}}}$ and $b'_{\mbox{\tiny{H}}}$ 
at four detectors. The readout recorded by these detectors 
is denoted by $(qrst)$. 
Since a polarizing beamsplitter transmits horizontal and reflects 
vertical polarization, the readout \lq\lq$q$'' refers to mode
$c'_{\mbox{\tiny{H}}}$, the readout \lq\lq$r$'' to mode
$c'_{\mbox{\tiny{V}}}$, the result \lq\lq$s$'' to mode
$b'_{\mbox{\tiny{V}}}$ 
and the readout \lq\lq$t$'' to mode $b'_{\mbox{\tiny{H}}}$. 
The range of values that can be assumed by these readouts 
depends on detector type and is clarified below.

The main task of this paper is to provide a model for 
an implementation of practical entanglement swapping. 
The goal is to model a realistic experiment with 
practical deficiencies. In a real-world scenario, the 
PDC sources are imperfect, creating not exactly one 
pair of entangled photons, but a superposition of 
alternatives that also includes the vacuum, independent 
pairs of photon-pairs, and higher pair-number contributions. 
This has been investigated before up to second order, see 
e.g.~\cite{Marcikic2002,Riedmatten2003}.
Furthermore, the detectors used to perform the 
Bell-state measurement are never perfect. 
They are usually inefficient to some degree, 
meaning that they sometimes do not detect 
existing photons. Detection inefficiencies 
are even further increased due to transmission losses 
between the source and detector. 
In theory, the latter can always be effectively 
taken into account by being included in the detector 
inefficiencies.
Moreover, the detectors 
are also subject to dark counts, meaning 
that they may click and indicate a detection event even if there are 
no photons incident into the detector. 
In this paper we will make a distinction between 
photon-counting detectors and detectors which cannot discriminate 
photon numbers. In each case, the recorded readout of 
a Bell-state measurement with inaccurate detectors 
will be denoted by $(qrst)$. In the first case,   
the measurement results  \lq\lq $q$'', 
\lq\lq $r$'',\lq\lq $s$'' and \lq\lq $t$'' 
can indicate any photon number $n\in\natuerl_0$, 
whereas in the second case they are records of 
yes/no events, namely either {\em \lq\lq at least one photon''} 
or {\em \lq\lq no photons''}. 

In what follows, we develop the basic ingredients of our theory.
We begin with a theoretical description of imperfect photon-pair sources. 
We proceed by providing a detector model that takes 
into account arbitrary detector inefficiencies as well as 
dark counts. Using a Bayesian reasoning approach we finally 
derive the resultant quantum state $\hat{\rho}^{qrst}$ of 
the remaining $a$ and $d$ modes in a realistic entanglement 
swapping experiment conditioned on the readout $(qrst)$ of 
an inaccurate Bell-state measurement.

\subsection{Modeling imperfect photon-pair sources}

Ideally, a photon-pair source would create exactly 
one entangled photon pair on demand. Such sources 
do not exist yet. Realistic sources are probabilistic  
generating photon pairs at random instances within 
those time intervals allowed by the (pulsed) pump laser,  
and occasionally emitting two or even more photon pairs, although the 
probability for higher order contributions is usually kept small. 
While other approaches exist, see e.g.~\cite{Duan2001}, 
parametric down conversion is the most common 
way to produce entangled photon pairs. 

In PDC, a crystal with an appreciably large $\chi^{(2)}$ 
nonlinearity is pumped by a laser field. Each of the pump photons 
can spontaneously decay into a pair of identical 
(degenerate PDC) or nonidentical photons (nondegenerate PDC). 
The rate of pair generation using PDC is proportional to the 
$\chi^{(2)}$ nonlinearity, the strength of the 
classical pump field and the interaction 
time. As shown in \cite{Barlett-et-al2001}, 
a PDC process can be described 
and mathematically represented by a one-parameter 
SU(1,1) transformation of the vacuum state:
\begin{equation}\label{SU(1,1)Trafo}
\Upsilon(\gamma)\vac=\exp\left(i\gamma\hat{K}_x\right)\vac\;,\quad\gamma\in\mathbb{R}\;.
\end{equation}
Here, $\hat{K}_x$ is one of the generators $\{\hat{K}_x,\hat{K}_y,\hat{K}_z\}$ 
of the SU(1,1) group defined by the commutator relations 
\begin{equation}\label{commutatorsGeneratorsFor-su(1,1)-Basis1}
\left[\hat{K}_x,\hat{K}_y\right]=-i\hat{K}_z\;,\quad
\left[\hat{K}_y,\hat{K}_z\right]=i\hat{K}_x\;,\quad 
\left[\hat{K}_z,\hat{K}_x\right]=i\hat{K}_y\;.
\end{equation}
For instance, in the case of type-I nondegenerate PDC, 
in which a pair of photons is created in the {\em same} polarization, 
and which we will consider throughout this paper, this generator is given by 
the following two-boson realization:
\begin{equation}\label{SU(1,1)TrafoGenerator}
\hat{K}_x=\frac{1}{2}\left(\hat{a}_{\mbox{\tiny{V}}}^{\dagger}
\hat{b}_{\mbox{\tiny{V}}}^{\dagger}+\hat{a}_{\mbox{\tiny{V}}}
\hat{b}_{\mbox{\tiny{V}}}\right) \;,
\end{equation}
where $\hat{a}_{\mbox{\tiny{V}}}$ and $\hat{b}_{\mbox{\tiny{V}}}$ 
are the annihilation operators corresponding to vertical 
polarizations of two different spatial modes $a$ and $b$.

As we can see, the resultant generated quantum state 
is not just a pair of photons, but a superposition of 
photon number states which particularly also includes 
the vacuum, pairs-of-pairs, and even higher order 
contributions. For small values of $\gamma$, the quantum 
state (\ref{SU(1,1)Trafo}) after a type-I nondegenerate PDC 
can be approximated as 
\begin{equation}\label{SU(1,1)TrafoSmallgamma}
\Upsilon(\gamma)\vac\approx
\vac+i\gamma\hat{K}_x\vac=
\vac+\frac{i\gamma}{2}\ket{0110}\;,
\end{equation}
where the Fock notation $\ket{ijkl}$ represents 
a state with $i,j,k,l$ photons in the $a_{\mbox{\tiny{H}}}$, 
$a_{\mbox{\tiny{V}}}$, 
$b_{\mbox{\tiny{V}}}$,  
$b_{\mbox{\tiny{H}}}$ modes, respectively. 
Please become aware of the chosen order \lq\lq$HVVH$''.  
This convention will turn out to be 
convenient with regard to the description 
of entanglement swapping, as it coincides with
the order of labels of the corresponding detection events 
in the readout $(qrst)$ of the Bell-state measurement, 
cf.~Fig.~\ref{Fig:entswap2}. The role of the vacuum state in the 
superposition (\ref{SU(1,1)TrafoSmallgamma}) is to allow for the 
particular feature that the generation of the desired photon pair occurs at
random instances of time. To be more precise, a photon-pair emission happens 
to be random within those periods of time, during which a pump laser field 
propagates through the crystal. That is, when a pump pulse is
sent, it will either lead to down conversion or not. 
There is a high probability that PDC will
not take place at all. The strong vacuum component implies this. It is for sure, though,
that there cannot be down-converted photon-pair creations during time
intervals between two successive laser pulses, i.e.\ when there is no laser
field propagating through the crystal.
The randomness of photon-pair production can be decreased  by using a crystal
with a larger $\chi^{(2)}$ nonlinearity or stronger pump fields, but this also happens 
at the cost of increased probability of the emission 
of multi-pairs of photons, which is disadvantageous and to be avoided 
as far as possible. On the other hand, multi-pair-emission events can never be 
completely excluded. 

For entanglement swapping, two PDC sources are required. 
As introduced above, the two different spatial modes 
of the first and of the second PDC source are labeled by 
$a$ and $b$, and by  $c$ and $d$, respectively.
In addition, the photons emitted by each source can have two mutually exclusive polarizations. 
We label them by $H$ and $V$, corresponding 
to {\em horizontal} and  {\em vertical} polarizations. Moreover, 
any symmetrical superposition of horizontal and vertical polarizations 
is possible. They have to be regarded as 
quantum alternatives and taken into account coherently. 
In this paper, we assume a type-I nondegenerate PDC for both sources. 
Furthermore, we elaborate our theory for {\em polarization entanglement}.
The preparation of polarization-entangled photon pairs can be 
experimentally realized using a pair of identical crystals stacked together 
such that their axes are orthogonal to each other, 
whereas the pump laser is diagonally polarized. 
Such a combination of two crystals effectively creates a PDC source 
producing two-mode squeezed states of the form given by 
Eq.~(\ref{SU(1,1)Trafo}) in each of the two orthogonal polarizations, 
with generators as given by Eq.~(\ref{SU(1,1)TrafoGenerator}) and 
the equivalent form for the $H$ polarization.
The total quantum state prepared by two PDC sources of this kind 
is then mathematically represented as:
\begin{eqnarray}\label{StateAfterPDC}
\ket{\chi}&=&\exp\left[i\chi\left(\hat{a}_{\mbox{\tiny{H}}}^{\dagger}
\hat{b}_{\mbox{\tiny{H}}}^{\dagger}+\hat{a}_{\mbox{\tiny{H}}}
\hat{b}_{\mbox{\tiny{H}}}\right)\right]\otimes
\exp\left[i\chi\left(\hat{a}_{\mbox{\tiny{V}}}^{\dagger}
\hat{b}_{\mbox{\tiny{V}}}^{\dagger}+\hat{a}_{\mbox{\tiny{V}}}
\hat{b}_{\mbox{\tiny{V}}}\right)\right]\nonumber\\ 
&&\otimes
\exp\left[i\chi\left(\hat{c}_{\mbox{\tiny{H}}}^{\dagger}
\hat{d}_{\mbox{\tiny{H}}}^{\dagger}+\hat{c}_{\mbox{\tiny{H}}}
\hat{d}_{\mbox{\tiny{H}}}\right)\right]\otimes
\exp\left[i\chi\left(\hat{c}_{\mbox{\tiny{V}}}^{\dagger}
\hat{d}_{\mbox{\tiny{V}}}^{\dagger}+\hat{c}_{\mbox{\tiny{V}}}
\hat{d}_{\mbox{\tiny{V}}}\right)\right]\ket{\mbox{vac}}\;.
\end{eqnarray}
We parameterize the generated quantum state by 
$\chi=\gamma/2\in\mathbb{R}$. 
Since $\chi$ is usually much smaller than one, 
the value $\chi^2$ is the {\em photon-pair production rate} of the PDC source, 
sometimes also referred to as its {\em brightness}. 
We will also call $\chi$ the {\em efficiency} of the source.   
In this paper we assume the same efficiency for both PDC sources.  
Too small values of  $\chi$ lead to a strong vacuum 
contribution, so that most of the time the sources do not emit any photon pairs. 
As  the value of $\chi$ increases the pollution from higher 
down-conversions becomes more and more important, see also~\cite{Bussieres2008}.

For the purpose of doing quantum optical calculations, it is convenient to
express the state~(\ref{StateAfterPDC}) in a {\em normal-ordered} form. 
This can be done as follows. Following \cite{Ban1993,PuriBook}, 
given two independent bosonic modes $a$ and $b$, 
we may choose a different basis of generators as compared to the basis in 
(\ref{commutatorsGeneratorsFor-su(1,1)-Basis1}) in order to obtain 
a two-mode bosonic representation of the su(1,1) Lie algebra:   
\begin{equation}
\hat{K}_+:=\hat{a}^{\dagger}\hat{b}^{\dagger}\;,\quad
\hat{K}_-:=\hat{a}\hat{b}\;,\quad
\hat{K}_0:=\frac{1}{2}\left(\hat{a}^{\dagger}\hat{a}+\hat{b}^{\dagger}\hat{b} +1\right)\;.
\end{equation}
The new generator basis  $\{\hat{K}_0,\hat{K}_+,\hat{K}_-\}$ of the 
su(1,1) Lie algebra satisfies the following commutator relations: 
\begin{equation}
\left[K_-,K_+\right]= 2K_0\;,\quad [K_0,K_{\pm}]=\pm K_{\pm}\;.
\end{equation}
According to \cite{Ban1993,PuriBook}, the following 
normal-order decomposition formula holds for exponential functions 
of the generators of the su(1,1) Lie algebra: 
\begin{equation}
\exp\left[\alpha_+K_++\alpha_0K_0+\alpha_-K_-\right]=\exp\left[A_+K_+\right]\exp\left[\ln(A_0)K_0\right]\exp\left[A_-K_-\right]\;,
\end{equation}
where $A_0,A_\pm$ are given by:
\begin{eqnarray}
A_\pm&=&\frac{(\alpha_\pm/\theta)\sinh\theta}{\cosh\theta-(\alpha_0/2\theta)\sinh\theta}\;,\\
A_0&=&\left[\cosh\theta-(\alpha_0/2\theta)\sinh\theta\right]^{-2}\;,\\
\theta&=&\left[(\alpha_0/2)^2-\alpha_+\alpha_-\right]^{1/2}\;.
\end{eqnarray}
Using this  decomposition rule we can derive the following special case which 
we need for our purpose: 
\begin{eqnarray}
\exp\left[i\chi\left(\hat{a}^{\dagger}\hat{b}^{\dagger}+\hat{a}\hat{b}\right)\right]&=&
\exp\left[\phi(\chi)\hat{a}^{\dagger}\hat{b}^{\dagger}\right]\nonumber\\
&&\times\exp\left[\omega(\chi)\left(\hat{a}^{\dagger}\hat{a}+\hat{b}^{\dagger}\hat{b}+1\right)\right]
\nonumber\\
&&\times\exp\left[\phi(\chi)\hat{a}\hat{b}\right]\;,
\end{eqnarray}
where we introduced the definitions: 
\begin{eqnarray}
\phi(\chi)&:=& i\tanh\chi\;,\\
\omega(\chi)&:=&-\ln[\cosh\chi]\;.
\end{eqnarray}
Each of the four factors in Eq.~(\ref{StateAfterPDC}) is of this form and can be
decomposed in this way. By doing so and using the fact that creation and 
annihilation operators corresponding to different optical modes commute, we arrive at:
\begin{eqnarray}\label{StateAfterPDC-2}
\ket{\chi}&=&\exp\left[4\omega(\chi)\right]\exp\left[\phi(\chi)
\left(\hat{a}_{\mbox{\tiny{H}}}^{\dagger}\hat{b}_{\mbox{\tiny{H}}}^{\dagger}+
\hat{a}_{\mbox{\tiny{V}}}^{\dagger}
\hat{b}_{\mbox{\tiny{V}}}^{\dagger}+
\hat{c}_{\mbox{\tiny{H}}}^{\dagger}
\hat{d}_{\mbox{\tiny{H}}}^{\dagger}+
\hat{c}_{\mbox{\tiny{V}}}^{\dagger}
\hat{d}_{\mbox{\tiny{V}}}^{\dagger}\right)\right]\nonumber\\
&&
\times\exp\left[\omega(\chi)\left(
\hat{a}_{\mbox{\tiny{H}}}^{\dagger}\hat{a}_{\mbox{\tiny{H}}}+
\hat{a}_{\mbox{\tiny{V}}}^{\dagger}\hat{a}_{\mbox{\tiny{V}}}+
\hat{b}_{\mbox{\tiny{H}}}^{\dagger}\hat{b}_{\mbox{\tiny{H}}}+
\hat{b}_{\mbox{\tiny{V}}}^{\dagger}\hat{b}_{\mbox{\tiny{V}}}+
\hat{c}_{\mbox{\tiny{H}}}^{\dagger}\hat{c}_{\mbox{\tiny{H}}}+
\hat{c}_{\mbox{\tiny{V}}}^{\dagger}\hat{c}_{\mbox{\tiny{V}}}+
\hat{d}_{\mbox{\tiny{H}}}^{\dagger}\hat{d}_{\mbox{\tiny{H}}}+
\hat{d}_{\mbox{\tiny{V}}}^{\dagger}\hat{d}_{\mbox{\tiny{V}}}\right)\right]\nonumber\\
&&\times\exp\left[\phi(\chi)
\left(\hat{a}_{\mbox{\tiny{H}}}\hat{b}_{\mbox{\tiny{H}}}+
\hat{a}_{\mbox{\tiny{V}}}
\hat{b}_{\mbox{\tiny{V}}}+
\hat{c}_{\mbox{\tiny{H}}}
\hat{d}_{\mbox{\tiny{H}}}+
\hat{c}_{\mbox{\tiny{V}}}
\hat{d}_{\mbox{\tiny{V}}}\right)\right]\ket{\mbox{vac}}\;.
\end{eqnarray}
It is easy to see that the last two exponential factors acting on the vacuum
state leave the latter unchanged. We are thus left with: 
\begin{equation}\label{StateAfterPDC-NormalOrder}
\ket{\chi}=\exp\left[4\omega(\chi)\right]\exp\left[\phi(\chi)
\left(\hat{a}_{\mbox{\tiny{H}}}^{\dagger}\hat{b}_{\mbox{\tiny{H}}}^{\dagger}+
\hat{a}_{\mbox{\tiny{V}}}^{\dagger}
\hat{b}_{\mbox{\tiny{V}}}^{\dagger}+
\hat{c}_{\mbox{\tiny{H}}}^{\dagger}
\hat{d}_{\mbox{\tiny{H}}}^{\dagger}+
\hat{c}_{\mbox{\tiny{V}}}^{\dagger}
\hat{d}_{\mbox{\tiny{V}}}^{\dagger}\right)\right]\ket{\mbox{vac}}\;.
\end{equation}
This is the normal-ordered form of the quantum state generated 
by the two PDC sources.

\subsection{Modeling imperfect  detectors}

We now present our theory of detectors which we apply to 
practical entanglement swapping. We begin with the description 
of ideal photon-number {\em discriminating} detectors and then stepwise 
allow for practical deficiencies. As a first step we take into 
account detector inefficiencies disregarding dark counts. In a 
second step we provide a detector model which also includes 
dark counts. By this means we get a theoretical description 
of photon-number discriminating detectors that are inefficient 
and subject to dark counts. Eventually, we will also have to  
acquiesce to the fact that most of the detectors presently available 
in laboratories cannot discriminate photon numbers, but instead 
effectively measure whether there are no photons or at least 
one photon in a mode. These kind of detectors are referred to as 
{\em threshold detectors}~\cite{BarlettSanders2002}. 
Finally, we explain our {\em Bayesian inference} approach 
which enables us to calculate the resultant quantum state 
after entanglement swapping in a realistic situation 
with imperfect detectors given the knowledge of it 
in the hypothetical ideal-detector scenario. 

\subsubsection{Unit-efficiency photon-number discriminating detectors with no dark counts}
Ideally, we would like to have a {\em unit-efficiency}, 
{\em photon-counting} detector that exhibits {\em no 
dark counts}. Such a detector never clicks when 
there is vacuum and always clicks whenever there 
are photons present in a certain mode, and the strength 
of the click provides information about the number 
of photons. Following \cite{BarlettSanders2002}, we 
refer to such a detector as an ideal, photon-number 
{\em discriminating}  detector. It is mathematically represented 
by a photon-counting projection-valued measure (PVM):
\begin{equation}\label{PVM}
\left\{\hat{\Pi}_n=\ket{n}\bra{n}\;,\quad n=0,1,2,3,\dots\right\}
\end{equation}
with respect to the Fock state basis $\{\ket{n}\;,\;
    n\in\natuerl_0\}$ 
of a certain mode.

\subsubsection{Inefficient photon-number discriminating detectors with no dark
  counts}
\label{Sec:InefficientDetectors} 
In practice, however, detectors always have a non-unit 
efficiency, meaning that even if photons are incident 
into the detector it has a finite probability not 
to trigger a click event. Throughout the paper, 
we denote the efficiency of a detector by $\eta$,   
with $ 0\le\eta\le 1$, where $\eta=1$ means a 100\% efficiency. 
Following \cite{RohdeRalph2006}, we model a detector 
efficiency $\eta$ by preceding a perfect, unit-efficiency 
detector with a beamsplitter possessing the transmittance $\eta$. 
This is illustrated in Fig.~\ref{DetectorModel}, if we 
replace there the thermal state in the second beamsplitter port
by the vacuum state. If $\hat{\rho}_{\mbox{\tiny sig}}$ is the input 
quantum state of the signal mode and $\hat{U}_{\mbox{\tiny BS}}(\eta)$ 
represents the unitary evolution corresponding to the 
beamsplitter transformation in 
the Schr{\"o}dinger picture, then the probability to 
detect $q$ photons, where $q\in\natuerl_0$, is given by
\begin{equation}
p_{\eta}\left(q|\hat{\rho}_{\mbox{\tiny sig}}\right)
=\Tr_{\mbox{\tiny trans}}\left[\hat{\Pi}_q\Tr_{\mbox{\tiny refl}}\left[
\hat{U}_{\mbox{\tiny BS}}(\eta)
\left(\hat{\rho}_{\mbox{\tiny sig}}\otimes\vac\bra{\mbox{vac}}\right)
\hat{U}^\dagger_{\mbox{\tiny BS}}(\eta)
\right]\hat{\Pi}_q\right]\;,
\end{equation}
where we first trace over the reflected mode 
then apply the PVM (\ref{PVM}) and finally take the 
trace with respect to the transmitted mode 
incident upon the perfect detector. 
Hereby we made the agreement that \lq\lq reflection'' and \lq\lq
transmission'' refer to the signal mode. 

For the purpose of this article, it is particularly 
important to expand on the special case in which  
the signal state is a photon number Fock state: 
$\hat{\rho}_{\mbox{\tiny sig}}=\ket{i}\bra{i}$. 
Given this input state, the conditional probability 
to measure $q$ photons with a unit-efficiency detector 
would be $p_{\eta=1}(q|i)=\delta_{qi}$, whereas the  
conditional probability with an $\eta$-efficiency detector ($\eta\not=1$)
amounts to:
\begin{eqnarray}
p_{\eta}\left(q|i\right)
&=&
\Tr_{\mbox{\tiny trans}}\left[\hat{\Pi}_q\Tr_{\mbox{\tiny refl}}\left[
\hat{U}_{\mbox{\tiny BS}}(\eta)\left(
\ket{i}\bra{i}\otimes\vac\bra{\mbox{vac}}\right)
\hat{U}^\dagger_{\mbox{\tiny BS}}(\eta)
\right]\hat{\Pi}_q\right]\;\nonumber\\
&=&
\left\{  \begin{array}{cc}
{i \choose q}\eta^q(1-\eta)^{i-q}
\quad
& \mbox{if $i\ge q$} \\
0
\quad &\mbox{if $i< q$} 
\end{array}  \right.\;.
\end{eqnarray}
Thus, the conditional probability to detect  $q$ photons given 
that $i$ photons are incident upon a non-unit efficiency detector 
is a Bernoulli distribution, as intuitively expected. 
Clearly, in order to be able to detect $q$ photons, 
the number $i$ of incident photons must not be smaller than $q$, 
as we have not allowed for dark counts yet.

Before we continue, we would like to note that the detector efficiencies $\eta$ 
of our theory are intended to take into account 
as much as possible all inefficiencies 
of the experiment. In particular, photon transmission losses,  
e.g., in fibers, filters, and other optical elements preceding a 
detector, can always be included in an {\em effective} efficiency  
of the detector. Thus, the efficiencies $\eta$ of our model 
are to be understood as effective efficiencies comprising 
the proper intrinsic detector efficiencies as well as all kind of other 
losses.

\subsubsection{Inefficient photon-number discriminating detectors with dark counts}
\label{Sec:ThermalSourceModel} 
\begin{figure}[tb] 
 \center{\includegraphics[width=0.3\linewidth]{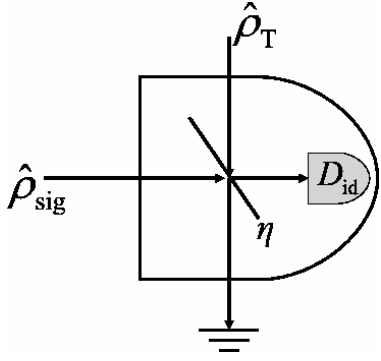}}
{\caption{A model for an 
imperfect detector with efficiency $\eta$ and dark counts
generated by a  fictitious thermal background source. The signal mode and 
the thermal mode represented by the quantum states 
$\hat{\rho}_{\mbox{\tiny sig}}$ and  $\hat{\rho}_{\mbox{\tiny T}}$, respectively, meet 
at a beamsplitter with transmittance $\eta$. One of its 
exits is directed to a perfect (ideal) detector ($D_{\mbox{\scriptsize id}}$), the photons of 
the second exit port are discarded. The perfect detector 
is assumed to be a {\em unit-efficiency} photon detector with {\em no 
dark counts}. We will make a distinction of two cases.
In the first instance we will assume the perfect detector 
to be photon-number discriminating. Later we consider the case 
where the perfect detector is a unit-efficiency threshold 
detector.\label{DetectorModel}}}
\end{figure}

We proceed by including the possibility of dark counts. 
We simulate dark counts by assuming the 
environment to be not in the 
vacuum state but in a thermal state of the form:
\begin{equation}\label{ThermalSource}
\hat{\rho}_{\mbox{\tiny\em T}}=\frac{1}{\cosh^2r}\sum_{n=0}^\infty \tanh^{2n}r\ket{n}\bra{n}\;.
 \end{equation}
This density operator models a thermal source with an average photon number
$\mbox{Tr}(\hat{\rho}_{\mbox{\tiny\em T}} \hat{n})=\sinh^2r$ and a pseudo temperature 
$T=\hbar\omega/(k_B\ln[\coth^2r])$. Thus, instead of assuming a vacuum state 
to be incident on the unused beamsplitter port, we combine our signal mode
which is to be measured with a thermal state (\ref{ThermalSource}) 
at the beamsplitter with transmissivity  $\eta$.
Our detector model is illustrated 
in Fig.~\ref{DetectorModel}.

The physical motivation behind this model is the association that the 
origin of dark counts stems from stray photons incident onto the detector from the 
environment. Furthermore, since the imperfect detector to be modeled has 
a non-unit efficiency, not all of the photons of the signal mode nor 
all photons from the thermal radiation cause a click in the detector, 
but just a fraction of them. This is effectively modeled by a 
beamsplitter whose transmittance $\eta$ is intended to represent the non-unit 
efficiency, followed by a subsequent perfect detector with {\em
  unit-efficiency} and {\em  no dark counts}.   
Let us stress, however, that the thermal radiation intended to be responsible 
for dark counts is entirely fictitious 
and does not have to correspond to a real, actually
existing photonic field. Any source of noise causing dark counts of 
the detector, e.g.\ electrical noise inside the detector, etc., 
can be effectively simulated by an imaginary 
thermal photonic field coupling into the detector from the 
environment via a beam splitter. This has been proven for homodyne 
detection by Appel et {\em al.} in~\cite{Appel2007}.
 
As will be shown in the appendix A, the probability for a dark count 
in a non-ideal threshold detector amounts to 
\begin{equation}\label{ProbabilityForDarkCount}
\wp_{\mbox{\tiny{dc}}}=\frac{(1-\eta)\tanh^2r}{1-\eta \tanh^2r}
=\frac{(1-\eta)\exp[-\frac{\hbar\omega}{kT}]}{1-\eta \exp[-\frac{\hbar\omega}{kT}]}\;.
\end{equation}
For any $0 \le \eta\le 1$, we can always find a pseudo-temperature 
$T=T(r)$ to model any arbitrary value of $\wp_{\mbox{\tiny{dc}}}$.
The case $\eta=1$ seems to exclude a non-vanishing
$\wp_{\mbox{\tiny{dc}}}\not=0$. However, we may take the simultaneous limit 
$\eta\rightarrow 1$ {\it and} $T\rightarrow\infty$ in such a way that 
any dark count probability is kept fixed.

Using this detector model, we can calculate the conditional 
probability to detect $q\in\natuerl_0$ photons  with a  photon-number discriminating detector 
possessing  arbitrary efficiency $\eta$ and any dark count 
probability $\wp_{\mbox{\tiny{dc}}}$, given an input 
quantum state $\hat{\rho}_{\mbox{\tiny sig}}$ of the signal 
mode, according to:
\begin{equation}\label{ResultForProb(q|rho)withDarkCounts}
p_{\eta,\wp_{\mbox{\tiny{dc}}}}(q|\hat{\rho}_{\mbox{\tiny sig}})
=\Tr_{\mbox{\tiny trans}}\left[\hat{\Pi}_q\Tr_{\mbox{\tiny refl}}\left[
\hat{U}_{\mbox{\tiny BS}}(\eta)
\left(\hat{\rho}_{\mbox{\tiny sig}}\otimes\hat{\rho}_{\mbox{\tiny\em T}}\right)
\hat{U}^\dagger_{\mbox{\tiny BS}}(\eta)
\right]\hat{\Pi}_q\right]\;.
\end{equation}
Again, we use the convention that \lq\lq reflection'' and \lq\lq
transmission'' refer to the signal mode. Furthermore, throughout the article, 
the subscripts $\eta$ and  $\wp_{\mbox{\tiny{dc}}}$ are used to express 
dependence on detector efficiency and dark count probability.
For the purpose of this paper, it is sufficient to know the conditional 
probabilities for the particular input quantum states 
$\hat{\rho}_{\mbox{\tiny sig}}=\ket{i}\bra{i}$, i.e.\ Fock states. 
We are therefore interested in  
$p_{\eta,\wp_{\mbox{\tiny{dc}}}}(q|i)$, with $i,q\in\natuerl_0$, 
which is the conditional probability to detect $q$ photons given 
that $i$ photons are incident upon a photon-number discriminating detector 
with efficiency $\eta$ and dark count probability $\wp_{\mbox{\tiny{dc}}}$. 
Please note that now, due to dark counts, $q$ may be greater than $i$. 
A derivation of these probabilities constitutes a significant technical 
part of the present work and can be found in the appendix B. The result reads:  
\begin{equation}\label{ResultForProb(q|i)}
p_{\eta,\wp_{\mbox{\tiny{dc}}}}(q|i)=
\left\{  \begin{array}{ll}
\frac{(1-\eta)(1-\wp_{\mbox{\tiny{dc}}})}{1-\eta(1-\wp_{\mbox{\tiny{dc}}})}
\left(\frac{\eta}{1-\eta}\right)^q
\left(1-\eta\right)^iG\left(i,q;\eta, \wp_{\mbox{\tiny{dc}}}\right)\quad
& \mbox{if $i\ge q$} \\
\frac{(1-\eta)(1-\wp_{\mbox{\tiny{dc}}})}{1-\eta(1-\wp_{\mbox{\tiny{dc}}})}
\left[\frac{1-\eta}{\eta}b(\eta,\wp_{\mbox{\tiny{dc}}})
\right]^{q-i}
\eta^iG(q,i;\eta, \wp_{\mbox{\tiny{dc}}})
\quad &\mbox{if $q\ge i$} \\
\end{array}  \right.\;\;,
\end{equation}
where 
\begin{equation}
\label{Eq:b-in-terms-of-pdc}
b(\eta,\wp_{\mbox{\tiny{dc}}}):=\eta\tanh^2r\equiv\left[ 1+\frac{1-\eta}{\eta
    \wp_{\mbox{\tiny{dc}}}}\right]^{-1}\;,
\end{equation}
and the function $G(\cdot,\cdot;\eta,
\wp_{\mbox{\tiny{dc}}}):\mathbb{N}_0^2\rightarrow \mathbb{R}$ is given as
follows. For $\kappa,\lambda\in \natuerl_0$, $\kappa\ge\lambda$, we define: 
\begin{equation}\label{Def:G-Function}
G(\kappa,\lambda;\eta, \wp_{\mbox{\tiny{dc}}}):=
\sum_{n=0}^\infty {\kappa \choose \lambda}{\kappa - \lambda +n \choose \kappa
  - \lambda}
\left[b(\eta,\wp_{\mbox{\tiny{dc}}})\right]^n\left[\, _2F_1\left(-n,-\lambda\,;\kappa - \lambda+1\,;\frac{\eta-1}{\eta}\right)\right]^2\;,
\end{equation} 
and $G(\kappa,\lambda;\eta, \wp_{\mbox{\tiny{dc}}}):=0$ in the case $ \kappa<\lambda$.
Here and in what follows, $_2F_1(\cdot,\cdot
,;\cdot\,;\cdot)$ denotes the Hypergeometric function
which is defined as 
\begin{eqnarray}
_2F_1(\alpha,\beta;\gamma;z)\label{Eq:DefHypergeometricFunction}
&:=& 1+\sum_{n=1}^\infty \frac{(\alpha)_n(\beta)_n}
{(\gamma)_n}\frac{z^n}{n!} \;\:,
\end{eqnarray}
where $(a)_n:=\Gamma(a+n)/\Gamma(a)$ is the Pochhammer symbol,
and $\Gamma(\cdot)$ is the Gamma function.
Please note that for $q=i$ the two results in Eq.~(\ref{ResultForProb(q|i)})
coincide.

\subsubsection{Inefficient threshold detectors with dark counts}

If we go a step further and acquiesce to the fact that photon-number 
discriminating detectors are a technological challenge and their 
realization is still in its infancy (however,
see~\cite{Rosenberg2005A,Rosenberg2005B}), 
then, to provide a description 
of entanglement swapping of the most practical relevance, we  
have to consider {\em threshold detectors}~\cite{BarlettSanders2002}. 
Ideally, such detectors effectively measure whether there are no photons 
or at least one photon in a mode. Following  \cite{BarlettSanders2002}, 
we refer to a unit-efficiency threshold detector with no dark counts as an 
{\em ideal threshold detector} (ITD). Thus, an ITD is mathematically 
described by the PVM
\begin{equation}\label{PVM-Threshold}
\left\{\hat{\Pi}_0=\ket{0}\bra{0}\;,\; 
\hat{\Pi}_{> 0}=\one-\hat{\Pi}_0\right\}\;.
\end{equation}
Inefficient threshold detectors with dark counts are  
contrived using the same detector model as above, but now 
with $D_{\mbox{\scriptsize id}}$ in Fig.~\ref{DetectorModel} being 
an ITD instead of an ideal photon-counting detector.
The relevant conditional probabilities are obtained in the same manner 
as above, using Eq.~(\ref{ResultForProb(q|rho)withDarkCounts}), but now with 
$q$ being either the event {\em \lq\lq no click''} or the complementary event 
{\em \lq\lq click''}, corresponding 
to the PVM elements $\hat{\Pi}_{0}$ or $\hat{\Pi}_{> 0}$, respectively. 
Again, for the purpose of entanglement swapping, we would like to 
know these conditional probabilities particularly for the signal input quantum states 
$\hat{\rho}_{\mbox{\tiny sig}}=\ket{i}\bra{i}$. The conditional probability of 
recording \lq\lq no click''  by a threshold detector 
with efficiency $\eta$ and dark count probability $\wp_{\mbox{\tiny{dc}}}$  
given that $i$ photons are incident upon it, can be directly obtained from 
the result (\ref{ResultForProb(q|i)}) by setting $q=0$. And the probability 
for the complementary event \lq\lq click'' is just one minus the latter 
probability: 
\begin{eqnarray}\label{ResultForProb(q|i)Threshold1}
p_{\eta,\wp_{\mbox{\tiny{dc}}}}(\mbox{\lq\lq no click"}|i)&=&
p_{\eta,\wp_{\mbox{\tiny{dc}}}}(q=0|i)\nonumber\\
&=&(1-\wp_{\mbox{\tiny{dc}}})
\big[1-\eta(1-\wp_{\mbox{\tiny{dc}}})\big]^i\\
p_{\eta,\wp_{\mbox{\tiny{dc}}}}(\mbox{\lq\lq click"}|i)
&=&1-p_{\eta,\wp_{\mbox{\tiny{dc}}}}(\mbox{\lq\lq no click"}|i)\nonumber\\
&=& 1-(1-\wp_{\mbox{\tiny{dc}}})
\big[1-\eta(1-\wp_{\mbox{\tiny{dc}}})\big]^i
\label{ResultForProb(q|i)Threshold2}
\end{eqnarray}

 \subsubsection{Bayesian updating based on evidence obtained by imperfect detectors}

In order to provide the resultant mixed quantum state 
of the remaining modes $a$ and $d$ depending on 
the result $(qrst)$ of a Bell-state measurement on the $c$ and
$b$ modes with imperfect detectors including the presence of dark counts,  
we proceed in the fashion of {\em Bayesian inference} 
and reasoning. We first assume the notional {\em ideal} situation that the detectors 
used for the Bell-state measurement are unit-efficiency photon-number discriminating 
detectors with no dark counts. In this hypothetical case we know how to calculate the 
probability for a certain measurement readout $(ijkl)$ of the 
perfectly accurate Bell-state measurement as well as the corresponding resultant 
pure quantum state, denoted by $\ket{\Phi_{ijkl}}$,  of the 
remaining modes $a$ and $d$ after the measurement,
using von Neumann's projection postulate \cite{vonNeumann1932}. 
We use this information as our hypothesis prior to observing 
evidence in the real experiment. The probability $p(ijkl)$ 
is our {\em prior probability} of the hypothesis that the 
resultant quantum state of the remaining modes $a$ and $d$
after the Bell measurement is given by $\ket{\Phi_{ijkl}}$. 

Before we proceed, we would like to make the following 
agreement. Throughout the paper we agree upon using the letters  
$q,r,s,t$ to denote the readouts of measurements 
using imperfect detectors, and the letters $i,j,k,l$ to label 
results of hypothetical measurements 
employing perfect photon-number discriminating detectors.
As for imperfect detectors, we differentiate between 
photon-number discriminating detectors and threshold detectors. 
In the first case $q,r,s,t\in \natuerl_0$, in the latter case 
$q,r,s,t\in\{\mbox{\em \lq\lq no click''},\mbox{\em \lq\lq click''}\}$.

Given an actual detector readout $(qrst)$ of an imperfect Bell 
measurement with inaccurate detectors including the presence 
of dark counts, we infer what an {\em ideal} four-tuple of detectors 
would have yielded, i.e., readout $(ijkl)$, with probability 
$P^{qrst}_{ijkl}:=p(ijkl|qrst)$. In the language of Bayesianism,
after the evidence $(qrst)$ has been observed by an imperfect Bell 
measurement, we update our knowledge with regard to the hypothesis 
according to {\em Bayes' theorem},
by means of which we can calculate the {\em posterior probability} 
of the hypothesis $(ijkl)$ given the obtained evidence $(qrst)$:
 \begin{eqnarray}\label{BayesTheorem2}
P^{qrst}_{ijkl}&\equiv& p(ijkl|qrst)
=\frac{p(qrst|ijkl)p(ijkl)}{p(qrst)}\nonumber\\ 
&=&\frac{p(qrst|ijkl)p(ijkl)}{\sum_{i',j',k',l'=0}^\infty 
p(qrst|i'j'k'l')p(i'j'k'l')}\;.
\end{eqnarray}
Here, $p(qrst|ijkl)$ indicates the conditional probability 
for obtaining the evidence $(qrst)$  
given that the hypothesis $(ijkl)$ would have happened 
to be true, if the detectors of our Bell-state measurement had been 
ideal. It is important to realize that $p(qrst|ijkl)$ is equivalent 
to the conditional probability of recording the readout $(qrst)$ 
given that exactly $i$, $j$, $k$ and $l$ photons are incident 
onto the four non-ideal detectors of the Bell-state measurement.
Hence, the resultant quantum state of the 
remaining modes $a$ and $d$ after recording the actual readout 
 $(qrst)$ at the inefficient detectors, is a {\em mixed state} of the form
\begin{equation}\label{StateAfterImperfectMeasurement}
\hat{\rho}^{qrst}= \sum_{ijkl}P^{qrst}_{ijkl}\ket{\Phi_{ijkl}}\bra{\Phi_{ijkl}}\;.
\end{equation}

In the next section we provide closed-form expressions for the conditional probabilities 
$P^{qrst}_{ijkl}$ depending on the experimental parameters $\chi$,
$\eta$ and $\wp_{\mbox{\tiny{dc}}}$. In the first instance we assume imperfect 
photon-number discriminating detectors, i.e.\ detectors that 
realize a photon-counting PVM but are in general inefficient 
($\eta\le 1$), and have a non-zero dark-count probability  $\wp_{\mbox{\tiny{dc}}}\not=0$.  
As our next step we consider imperfect threshold detectors.
Again, in the style of Bayesian reasoning we upgrade our probabilities 
of the density operator (\ref{StateAfterImperfectMeasurement}) 
using the evidence that the measurement outcomes $q,r,s,t$ 
of a Bell-state measurement with threshold detectors 
can be either {\em \lq\lq no click''} or  {\em \lq\lq click''}. 
The resultant quantum state obtained in this way is the 
most relevant result, as it refers to the 
most common practical situation, namely entanglement swapping 
using ordinary inaccurate threshold detectors.

\subsection{Quantum state after entanglement swapping}
\noindent

Using the tools presented in the previous sections we are 
now in a position to derive the resultant quantum state (\ref{StateAfterImperfectMeasurement}) 
after a realistic entanglement swapping with imperfect sources 
and imperfect detectors. 

We start with the quantum state provided by the two PDC sources, 
Eq.~(\ref{StateAfterPDC-NormalOrder}). 
Suppose our four detectors of the Bell measurement on the modes 
$b$ and $c$  were perfect, i.e., had a unit-efficiency 
$\eta=1$ and no detector dark counts. Then, to give the ideal readout
$(ijkl)$, after applying the balanced beamsplitter transformation 
$B=\frac{1}{\sqrt{2}}\left({\;\,1\atop -1}{1\atop 1}\right)$ to 
 modes $b$ and $c$ using the rule \cite{VogelWelsch2006}
\begin{equation}
{\hat{b}_{\mbox{\tiny{H}}}\choose \hat{c}_{\mbox{\tiny{H}}}}\rightarrow B^\dagger{\hat{b'}_{\mbox{\tiny{H}}}\choose
  \hat{c'}_{\mbox{\tiny{H}}}}\quad,\quad
{\hat{b}^\dagger_{\mbox{\tiny{H}}}\choose \hat{c}^\dagger_{\mbox{\tiny{H}}}}\rightarrow B^T{\hat{b'}^\dagger_{\mbox{\tiny{H}}}\choose
  \hat{c'}^\dagger_{\mbox{\tiny{H}}}}\;,
\end{equation}
and similarly for the vertical polarization,
the resulting four-mode quantum state $\hat{U}_B\ket{\chi}$ 
is to be projected onto the subspace corresponding to 
the projector 
\begin{equation}\label{Projector-ijkl}
\hat{\Pi}^{(ijkl)}_{c'_H,c'_V,b'_V,b'_H}:=\left(\ket{i}\bra{i}\right)_{c'_H}\otimes(\ket{j}\bra{j})_{c'_V}
\otimes
(\ket{k}\bra{k})_{b'_V}\otimes(\ket{l}\bra{l})_{b'_H}\otimes\one_{a_H}
\otimes\one_{a_V}\otimes\one_{d_V}\otimes\one_{d_H}\;.
\end{equation}
The modes $c'_H,c'_V,b'_V$ and $b'_H$ are the output modes of the
  balanced beamsplitter, cf.~Fig.~\ref{Fig:entswap2}.
The operator $\hat{U}_B$ represents the unitary evolution corresponding to the 
beamsplitter transformation in the Schr{\"o}dinger picture, which we use. 
Here and in what follows, $\ket{n}$ denotes an $n$-photon Fock state. It should be clear
from the context to which mode it refers. Accordingly,
$(\ket{i}\bra{i})_{c'_H}$ represents a projection operator corresponding to the 
Fock state $\ket{i}$ of mode $c'_H$, etc.. 
The projection (\ref{Projector-ijkl}) followed by state normalization 
yields the following post-measurement quantum state :
\begin{equation}\label{StateAfterMeasurementIncluding-cb}
\frac{\hat{\Pi}^{(ijkl)}_{c'_H,c'_V,b'_V,b'_H}\hat{U}_B\ket{\chi}}{
\left\|\hat{\Pi}^{(ijkl)}_{c'_H,c'_V,b'_V,b'_H}\hat{U}_B\ket{\chi} \right\|}=
\ket{i^{c'_H}\,j^{c'_V}\,k^{b'_V}\,l^{b'_H}}\otimes\ket{\Phi_{ijkl}}\;,
\end{equation}
with the first factor  
being the Fock state with $i,j,k,l$ photons in the modes $c'_H$, $c'_V$,  
$b'_V$, and $b'_H$, respectively, and 
\begin{eqnarray}\label{Result-For:Phi(ijkl)}
\ket{\Phi_{ijkl}}
&\equiv&\frac{1}{\sqrt{i!j!k!l!}}
\left(\frac{\hat{d}_{\mbox{\tiny{H}}}^{\dagger}-\hat{a}_{\mbox{\tiny{H}}}^{\dagger}}{\sqrt{2}}
\right)^i
\left(\frac{\hat{d}_{\mbox{\tiny{V}}}^{\dagger}-\hat{a}_{\mbox{\tiny{V}}}^{\dagger}}{\sqrt{2}}
\right)^j
\left(\frac{\hat{a}_{\mbox{\tiny{V}}}^{\dagger}+\hat{d}_{\mbox{\tiny{V}}}^{\dagger}}{\sqrt{2}}
\right)^k
\left(\frac{\hat{a}_{\mbox{\tiny{H}}}^{\dagger}+\hat{d}_{\mbox{\tiny{H}}}^{\dagger}}{\sqrt{2}}
\right)^l\ket{\mbox{vac}}\nonumber\\
&=&\frac{1}{(\sqrt{2})^{i+j+k+l}\sqrt{i!j!k!l!}}
\sum_{\mu=0}^i
\sum_{\nu=0}^j
\sum_{\kappa=0}^k
\sum_{\lambda=0}^l
(-1)^{\mu+\nu}
{i \choose \mu}
{j \choose \nu}
{k \choose \kappa}
{l \choose \lambda}
\\ \nonumber
&& \times \left(\hat{a}_{\mbox{\tiny{H}}}^{\dagger}\right)^{\mu+\lambda}
\left(\hat{a}_{\mbox{\tiny{V}}}^{\dagger}\right)^{\nu+\kappa}
\left(\hat{d}_{\mbox{\tiny{H}}}^{\dagger}\right)^{i+l-\mu-\lambda}
\left(\hat{d}_{\mbox{\tiny{V}}}^{\dagger}\right)^{j+k-\nu-\kappa}
\ket{\mbox{vac}}\;.
\end{eqnarray}
Since the photons of the $c$ and $b$ modes are destroyed in the measurement 
process, we discard the first factor of 
Eq.~(\ref{StateAfterMeasurementIncluding-cb}). The second factor, 
$\ket{\Phi_{ijkl}}$, is the resultant pure state 
of the remaining $a$ and $d$ modes:
The corresponding probability of the hypothetical ideal measurement readout $(ijkl)$ is given by:
\begin{equation}\label{Result:p(ijkl)}
p(ijkl)=\,\left\| \hat{\Pi}^{(ijkl)}_{c'_H,c'_V,b'_V,b'_H}\hat{U}_B\ket{\chi}
\right\|^2=\frac{\big[\tanh\chi\big]^{2(i+j+k+l)}}{\cosh^8\chi}\;.
\end{equation}

In an actual experiment, however, detectors have an efficiency 
which is appreciably less than 100\% and in addition 
exhibit dark counts. Given an actual detector readout $(qrst)$ 
of an imperfect Bell measurement with faulty photon-number discriminating 
detectors characterized by efficiency $\eta<1$ and dark count probability $\wp_{\mbox{\tiny{dc}}}$, 
we do not know the corresponding resultant post-measurement pure quantum state 
for the remaining $a$ and $d$ 
modes nor the probability of its occurrence. We can, however, calculate 
the posterior conditional probability for any readout $(ijkl)$ an ideal 
four-tuple of detectors would have yielded, i.e., the probability 
$P^{qrst}_{ijkl}(\eta,\wp_{\mbox{\tiny{dc}}}) \equiv 
p_{\eta,\wp_{\mbox{\tiny{dc}}}}(ijkl|qrst)$. As explained in the previous section,  
this is done using Bayes' theorem (\ref{BayesTheorem2}), according to 
which we achieve our goal if we know the conditional probabilities  
$p_{\eta,\wp_{\mbox{\tiny{dc}}}}(qrst|ijkl)$ for all possible events $(ijkl)\in\natuerl_0^4$.
Since the four detectors are {\em statistically 
independent} from one another, these probabilities factorize into four terms:  
\begin{equation}\label{FactorizationConditionalProbability}
p_{\eta,\wp_{\mbox{\tiny{dc}}}}(qrst|ijkl)=p_{\eta,\wp_{\mbox{\tiny{dc}}}}(q|i)\,p_{\eta,\wp_{\mbox{\tiny{dc}}}}(r|j)\,
p_{\eta,\wp_{\mbox{\tiny{dc}}}}(s|k)\,p_{\eta,\wp_{\mbox{\tiny{dc}}}}(t|l)\;, 
\end{equation}
where each of these factors is given by the expression (\ref{ResultForProb(q|i)}).
The factorization of the conditional probabilities together with 
result (\ref{Result:p(ijkl)}) imply  
a factorization of the posterior probabilities 
of Eq.~(\ref{BayesTheorem2}):
\begin{equation}\label{AccordingToBayesTheorem}
P^{qrst}_{ijkl}(\chi,\eta,\wp_{\mbox{\tiny{dc}}}) 
=f^q_i(\chi,\eta, \wp_{\mbox{\tiny{dc}}})
f^r_j(\chi,\eta, \wp_{\mbox{\tiny{dc}}})
f^s_k(\chi,\eta, \wp_{\mbox{\tiny{dc}}})
f^t_l(\chi,\eta, \wp_{\mbox{\tiny{dc}}})\;, 
\end{equation}
where we now explicitely express their dependence on the 
experimental parameters $\chi,\eta$ and $ \wp_{\mbox{\tiny{dc}}}$. 
By inserting result (\ref{Result:p(ijkl)}) and our physical 
assumption  (\ref{FactorizationConditionalProbability}) into 
(\ref{BayesTheorem2}), we have:
\begin{equation}\label{Def:f^q_i}
f^q_i(\chi,\eta,\wp_{\mbox{\tiny{dc}}})\equiv
\frac{p_{\eta,\wp_{\mbox{\tiny{dc}}}}(q|i)\tanh^{2i}\chi}{
\sum_{i'=0}^\infty  p_{\eta,\wp_{\mbox{\tiny{dc}}}}(q|i')\tanh^{2i'}(\chi)}
\end{equation}
and equivalent formulae for $f^r_j(\chi,\eta, \wp_{\mbox{\tiny{dc}}})$, 
$f^s_k(\chi,\eta, \wp_{\mbox{\tiny{dc}}})$ and $f^t_l(\chi,\eta,
\wp_{\mbox{\tiny{dc}}})$.
\vspace{1mm}
By inserting (\ref{ResultForProb(q|i)}) into (\ref{Def:f^q_i})  
we finally obtain  the following exact closed-form 
solution: 
\begin{equation}\label{ResultFor:f^q_i} 
f^q_i(\chi,\eta,\wp_{\mbox{\tiny{dc}}})= 
\left\{  \begin{array}{ll}
\frac{\tanh^{2i}(\chi)(1-\eta)^{i-q}
G(i,q;\eta, \wp_{\mbox{\tiny{dc}}})}{
g(q;\chi,\eta,\wp_{\mbox{\tiny{dc}}})}\quad
& \mbox{if $i\ge q$} 
\\
\frac{\tanh^{2i}(\chi)\eta^{2(i-q)}
(1-\eta)^{q-i}b^{q-i}(\eta,\wp_{\mbox{\tiny{dc}}})
G(q,i;\eta, \wp_{\mbox{\tiny{dc}}})}{
g(q;\chi,\eta,\wp_{\mbox{\tiny{dc}}})}
\quad &\mbox{if $q\ge i$} 
\\
\end{array}  \right.
\end{equation}
with the common denominator 
\begin{eqnarray}\label{DenominatorOfResultFor:f^q_i}
g(q;\chi,\eta,\wp_{\mbox{\tiny{dc}}})&:=&
\sum_{i'=0}^q\tanh^{2i'}(\chi)\,\eta^{2(i'-q)}(1-\eta)^{q-i'}
b^{q-i'}(\eta,\wp_{\mbox{\tiny{dc}}})G(q,i';\eta, \wp_{\mbox{\tiny{dc}}})\nonumber \\
&&+\sum_{i'=q+1}^\infty\tanh^{2i'}(\chi)(1-\eta)^{i'-q}
G(i',q;\eta, \wp_{\mbox{\tiny{dc}}})
\end{eqnarray}
\vspace{1mm}
\noindent
and similar expressions for $f^r_j(\chi,\eta, \wp_{\mbox{\tiny{dc}}})$,
$f^s_k(\chi,\eta, \wp_{\mbox{\tiny{dc}}})$ and $f^t_l(\chi,\eta,
\wp_{\mbox{\tiny{dc}}})$.
Eqs.~(\ref{AccordingToBayesTheorem}),(\ref{ResultFor:f^q_i}) 
and (\ref{DenominatorOfResultFor:f^q_i}) form our result 
for {\em photon-number discriminating} detectors. Knowing the 
probabilities $P^{qrst}_{ijkl}(\chi,\eta,\wp_{\mbox{\tiny{dc}}})$ 
for all possible hypothetical readouts $(ijkl)$ with $i,j,k,l\in\mathbb{N}_0$    
allows computation of the mixed quantum state (\ref{StateAfterImperfectMeasurement}) 
for the remaining $a$ and $d$ modes after an imperfect Bell measurement 
with measurement result $(qrst)$ using photon-number discriminating
detectors with efficiency $\eta$ and dark count probability
$\wp_{\mbox{\tiny{dc}}}$.

Let us consider the special case when dark counts 
are absent, i.e. $\wp_{\mbox{\tiny{dc}}}=0$. With
Eq.~(\ref{ProbabilityForDarkCount}) this leads to
$(1-\eta)\tanh^2r=0$ and therefore $\tanh^2r=0$ as $\eta\not=1$ 
in general. This in turn implies  
$b(\eta,\wp_{\mbox{\tiny{dc}}}=0)=\eta\tanh^2r\equiv 0$  
as $\eta\not=0$, 
and as a consequence also $G(\kappa,\lambda;\eta, \wp_{\mbox{\tiny{dc}}}=0)={ \kappa \choose \lambda}$. 
According to our general result
(\ref{ResultFor:f^q_i}) the absence of dark counts,
$\wp_{\mbox{\tiny{dc}}}=0$, implies  $i\ge q$, again as $b(\eta,\wp_{\mbox{\tiny{dc}}}=0)$ 
must be zero, and we obtain: 
\begin{eqnarray}
f^q_i(\chi,\eta,\wp_{\mbox{\tiny{dc}}}=0)&=&
\frac{{ i \choose q}(1-\eta)^{i-q}
\tanh^{2i}\chi}{
\sum_{i'=q}^\infty{ i' \choose q}(1-\eta)^{i'-q}
\tanh^{2i'}\chi}\nonumber\\
&=&
\frac{{ i \choose  q}(1-\eta)^{i-q}\big[\tanh^2\chi\big]^{i-q}}
{\sum_{i"=0}^\infty{ q+i" \choose
  q}\big[(1-\eta)\tanh^{2}\chi\big]^{i"}}
\nonumber\\
&=&
\frac{{ i \choose q}\big[\tanh^2\chi\big]^{i-q}(1-\eta)^{i-q}}
{\frac{1}{\Big(1-\big[(1-\eta)\tanh^{2}\chi\big]\Big)^{q+1}}}\;,
\end{eqnarray} 
i.e., we get the result: 
\begin{equation}
f^q_i(\chi,\eta,\wp_{\mbox{\tiny{dc}}}=0)=
{ i \choose  q}
\big[(1-\eta)\tanh^2\chi\big]^{i-q}\Big(1-\big[(1-\eta)\tanh^{2}\chi\big]\Big)^{q+1}\;.
\end{equation}
If we also let the efficiency $\eta$ go to unity, $\eta\rightarrow 1$, we arrive at: 
\begin{equation}\label{f^q_iIdealDet}
\lim_{\eta\rightarrow 1}f^q_i(\chi,\eta,\wp_{\mbox{\tiny{dc}}}=0)
=\delta_{qi}\;,
\end{equation}
where $\delta_{qi}$ is the Kronecker delta symbol.
This is also what one should expect in the ideal case of perfect detectors 
with no dark counts. Please note, that with $\eta\rightarrow 1$ we have also
  implicitly assumed that there are no transmission losses between the 
sources and detectors.

Acquiescing to the fact that photon-number discriminating detectors are still 
a technological challenge, we now consider the more 
practical situation of threshold detectors, in which case 
the events $(qrst)$ of the Bell measurement consist of 
readouts $q,r,s,t\in\{\mbox{\em \lq\lq no click''},\mbox{\em \lq\lq click''}\}$.
The conditional probabilities 
are given by 
(\ref{ResultForProb(q|i)Threshold1}) and 
(\ref{ResultForProb(q|i)Threshold2}).
Calculation of the functions $f^q_i(\chi,\eta,\wp_{\mbox{\tiny{dc}}})$ in  
Eq.~(\ref{Def:f^q_i})  yields the following result: 
\begin{eqnarray}\label{ResultFor:f^q_iForthresholddetectors1}
f^{\mbox{\tiny \lq\lq no click"}}_i(\chi,\eta,\wp_{\mbox{\tiny{dc}}})&=&
\big[h(\chi,\eta,\wp_{\mbox{\tiny{dc}}})\big]^i
\big(1-h(\chi,\eta,\wp_{\mbox{\tiny{dc}}})\big)\;,\\
f^{\mbox{\tiny \lq\lq click"}}_i(\chi,\eta,\wp_{\mbox{\tiny{dc}}})
&=&\frac{
\Big\{\tanh^{2i}\chi-(1-\wp_{\mbox{\tiny{dc}}})\left[h(\chi,\eta,\wp_{\mbox{\tiny{dc}}})
\right]^i\Big\}}{
\cosh^2\chi-\frac{1-\wp_{\mbox{\tiny{dc}}}}{1-h(\chi,\eta,\wp_{\mbox{\tiny{dc}}})}}\;,
\label{ResultFor:f^q_iForthresholddetectors2}
\end{eqnarray}
where we have introduced the definition:
\begin{equation}\label{ResultFor:f^q_iForthresholddetectorsDef-h}
h(\chi,\eta,\wp_{\mbox{\tiny{dc}}}):= \left[1-\eta(1-\wp_{\mbox{\tiny{dc}}})\right]\tanh^{2}\chi\;.
\end{equation}
Eq.~(\ref{AccordingToBayesTheorem}) together with
(\ref{ResultFor:f^q_iForthresholddetectors1} -\ref{ResultFor:f^q_iForthresholddetectorsDef-h}) 
form our result for threshold detectors. Please note, that in the ideal case, $\eta= 1$ and
$\wp_{\mbox{\tiny{dc}}}= 0$, our result reduces to what we should expect:   
\begin{equation}
f^{\mbox{\tiny \lq\lq no click"}}_i(\chi,\eta=1,\wp_{\mbox{\tiny{dc}}}=0)
=\delta_{i,0}
\end{equation}
and
\begin{eqnarray}
f^{\mbox{\tiny \lq\lq click"}}_i(\chi,\eta=1,\wp_{\mbox{\tiny{dc}}}=0)
&=&\left\{  \begin{array}{ll}
0 & \quad\mbox{if $i=0$} \\
\frac{\tanh^{2i}\chi}{\sinh^2\chi} &\quad \mbox{if $i\ge 1$} \\
\end{array}  \right.\\
\sum_{i=1}^\infty f^{\mbox{\tiny \lq\lq
    click"}}_i(\chi,\eta=1,\wp_{\mbox{\tiny{dc}}}=0)&=&
\frac{\tanh^{2}\chi}{\sinh^2\chi}\sum_{i'=0}^\infty
\tanh^{2i'}\chi=1\;.
\end{eqnarray}

As for the peculiar non-relevant marginal situation $\eta= 0$ and
$\wp_{\mbox{\tiny{dc}}}= 0$, we note that in this case 
there can be no click events, so that 
$f^{\mbox{\tiny \lq\lq click"}}_i(\chi,\eta=0,\wp_{\mbox{\tiny{dc}}}=0)$ 
becomes a meaningless conditional probability. So it is not a flaw that 
the latter is not defined in this marginal situation. On the other 
hand, we get $ f^{\mbox{\tiny \lq\lq no
    click"}}_i(\chi,\eta=0,\wp_{\mbox{\tiny{dc}}}=0)=\tanh^{2i}\chi/\cosh^2\chi$ 
and $ \sum_{i=0}^\infty f^{\mbox{\tiny \lq\lq no
    click"}}_i(\chi,\eta=0,\wp_{\mbox{\tiny{dc}}}=0)=1$.

Our derivation of the resultant quantum state (\ref{StateAfterImperfectMeasurement}) 
after an imperfect entanglement swapping is based on Bayesian  
reasoning and the physical assumption that the four detectors of the 
Bell-state measurement are statistically 
independent from one another implying the property of 
Eq.~(\ref{FactorizationConditionalProbability}).
The latter property involves a factorization of the probabilities 
$P^{qrst}_{ijkl}$ into four terms corresponding to the 
four independent detectors, cf.\ Eq.~(\ref{AccordingToBayesTheorem}). 
An obvious generalization of our results is to allow different 
efficiencies as well as different dark count probabilities for the 
four detectors. The derivation is very similar and the result reads:
\begin{equation}\label{AccordingToBayesTheoremGeneralized}
P^{qrst}_{ijkl}(\,\chi,\{\eta_{\nu}\},
\{{\wp_{\mbox{\tiny{dc}}}}_{\nu}\}) 
=f^q_i(\chi,\eta_1, {\wp_{\mbox{\tiny{dc}}}}_1)
f^r_j(\chi,\eta_2, {\wp_{\mbox{\tiny{dc}}}}_2)
f^s_k(\chi,\eta_3, {\wp_{\mbox{\tiny{dc}}}}_3)
f^t_l(\chi,\eta_4, {\wp_{\mbox{\tiny{dc}}}}_4)\;, 
\end{equation}
where $\eta_{\nu}$ and ${\wp_{\mbox{\tiny{dc}}}}_{\nu}$,
$\nu=1,2,3,4$, denote the arbitrary and in general 
different efficiencies and dark count probabilities 
of the four detectors. The functions 
$f^m_n(\chi,\eta_{\nu}, {\wp_{\mbox{\tiny{dc}}}}_{\nu})$ 
are given either by the result (\ref{ResultFor:f^q_i}) in the case of 
photon-number discriminating detectors or 
by
(\ref{ResultFor:f^q_iForthresholddetectors1},\ref{ResultFor:f^q_iForthresholddetectors2})
in the case of threshold detectors.

\section{Comparison with experimental entanglement swapping}
\label{ComparisonWithExperimentalSwapping}

To evaluate the value of our model we test it against 
results of real entanglement swapping experiments~\cite{PanBouwmeesterWeinfurterZeilinger1998,
JenneweinWeihsPanZeilinger2002,RiedmattenMarcikicHouwelingenTittelZbindenGisin2005}.
In these experiments entanglement verification is accomplished either by 
observing the visibility of four-fold coincidences of four threshold
detectors (two for the Bell-state measurement and two for monitoring 
the modes $a$ and $d$, one on each side), obtained via 
variable polarization directions of analyzers, or by measuring 
certain correlation coefficients for polarization related to tests 
of the CHSH Bell inequality~\cite{Clauser1969}. Assuming the 
resultant quantum state of the photons of the remaining modes $a$ and $d$ 
to be given by a Werner state, cf.~\cite{Tittel2003}, 
\begin{equation}
\label{Werner-state}
\hat{\rho}^{qrst}=F\ket{\psi^T}\bra{\psi^T}+\frac{1-F}{3}\left(\one-\ket{\psi^T}\bra{\psi^T}\right)\;,
\end{equation}
where $\ket{\psi^T}$ is some target Bell state, 
the visibility is directly connected to the fidelity 
$F=\bra{\psi^T}\hat{\rho}^{qrst}\ket{\psi^T}$ 
via the relation (cf.~\cite{RiedmattenMarcikicHouwelingenTittelZbindenGisin2005}):
\begin{equation}
\label{F-V-Relation}
V=(4F-1)/3\;.
\end{equation}
Let us emphasize, however, that the assumption (\ref{Werner-state}) 
is justified only for the post-selected quantum state 
$ \hat{\rho}^{qrst}_{\mbox{\tiny
    postsel}}$, i.e, provided 
that click events are observed in both the $a$ and $d$ modes.  
This issue will be clarified below. 
The visibility is obtained according to $V=\left(\mbox{Max}-\mbox{Min}\right)/
\left(\mbox{Max}+\mbox{Min}\right)$, where \lq\lq $\mbox{Max}$'' and \lq\lq $\mbox{Min}$'' 
denote the maximal and minimal values of 
the four-fold coincidence rate as a function of some polarization 
angle, respectively. The relation between visibility and 
the correlation coefficient $S$ of the CHSH Bell inequality is 
$S=2\sqrt{2}\,V$, cf.~\cite{Rarity1990}.

To simulate a four-fold coincidence experiment and provide predictions 
for the visibility we proceed as follows. In accordance with the 
experimental situation of references~\cite{PanBouwmeesterWeinfurterZeilinger1998,
RiedmattenMarcikicHouwelingenTittelZbindenGisin2005} 
we choose to consider four-fold coincidence events in which the result 
of the Bell measurement on the $b$ and $c$ modes corresponds --- in the ideal 
case scenario --- to a projection onto the Bell state 
\begin{equation}
\ket{\Psi^-}_{cb}=\frac{1}{\sqrt{2}}\left(\hat{c}_{\mbox{\tiny{H}}}^{\dagger} \hat{b}_{\mbox{\tiny{V}}}^{\dagger}-
\hat{c}_{\mbox{\tiny{V}}}^{\dagger} \hat{b}_{\mbox{\tiny{H}}}^{\dagger}\right)\ket{\mbox{vac}}\;.
\end{equation}
According to the common argument on entanglement swapping,   
this would mean that the remaining modes $a$ and $d$ are left in the 
same Bell state, i.e.\  $\ket{\Psi^-}_{ad}$.
To verify this entangled state by measurements 
on the $a$ and $d$ modes is the objective of the experiment of 
  \cite{PanBouwmeesterWeinfurterZeilinger1998}. The experimental 
situations in references
\cite{JenneweinWeihsPanZeilinger2002,RiedmattenMarcikicHouwelingenTittelZbindenGisin2005} 
are very similar. 

Let us for a moment assume that our detectors and sources are ideal.
Then, a projection onto the Bell state $\ket{\Psi^-}_{cb}$ is achieved 
whenever there are coincidence clicks of the two detectors for the $c'_H$ 
and the $b'_V$ mode or, vice versa, of the two detectors for the $c'_V$ and 
$b'_H$ modes. 
To understand this issue, we can exploit the fact that 
the Bell state $\ket{\Psi^-}_{cb}$ is antisymmetric under exchange 
of labels $c$ and $b$ implying fermionic statistics in the 
spatial behavior of the two photons, which means that they 
have to emerge from different ports of the balanced beam-splitter  
(cf.~Fig.~\ref{Fig:entswap2}). The other three Bell states are 
symmetric with respect to exchange of labels $c$ and $b$ involving 
bosonic statistics in the spatial behavior, meaning that the photons 
will emerge at the same output port of the beamsplitter. Hence, 
in the ideal case of perfect sources and detectors, observing coincidences 
between two detectors on both sides of the beamsplitter is an experimental 
evidence for a projection onto the state $\ket{\Psi^-}_{cb}$ and thus 
also a preparation of the state $\ket{\Psi^-}_{ad}$ for the $a$ and $d$ 
modes. Only the coincidences of \lq\lq $c'_H$ and $b'_V$'' or \lq\lq $c'_V$ and 
$b'_H$'' clicks are possible, but not of two clicks corresponding to 
the same polarization, i.e.,  \lq\lq $c'_H$ and $b'_H$'' 
or \lq\lq $c'_V$ and $b'_V$'', since the polarizations in the Bell state 
$\ket{\Psi^-}_{cb}$ are anti-correlated. 

Yet, in a real experiment 
scenario the PDC sources and detectors are not ideal. In particular, 
we have to 
allow for the rare but non-negligible faulty events due to emission of 
multi-photon pairs as well as dark counts, and also take into account 
transmission losses and detector inefficiencies,
so that the actual quantum state of the remaining $a$ and $d$ modes 
after the suggested Bell measurement will deviate from $\ket{\Psi^-}_{ad}$. 
Within the setting of our model with imperfect threshold detectors, 
a non-ideal projection onto the Bell state
$\ket{\Psi^-}_{cb}$ is achieved whenever one of the following Bell measurement 
events $(q,r,s,t)$ is obtained: $(1,0,1,0)$ or $(0,1,0,1)$. Here and in what
follows, $q=1$ means a {\em click} and $q=0$ that the detector {\em does not
click}. 

Before we proceed we would like to comment on the following 
interesting observation. Even if the detectors of the Bell-state 
measurement were ideal ($\eta=1,\wp_{\mbox{\tiny{dc}}}=0$), 
they would never indicate a projection onto $\ket{\Psi^-}_{cb}$ and 
as a consequence prepare the Bell state  $\ket{\Psi^-}_{ad}$  
unless also the photon-pair sources were ideal.  
The multi-pair nature of the PDC sources precludes 
a projection onto the Bell state $\ket{\Psi^-}_{cb}$ 
regardless of the quality of the detectors used for 
the  Bell-state measurement. To understand this feature, 
let us consider the outcome $(1010)$ of the Bell-state measurement 
on the $b$ and $c$ modes and assume ideal photon-number discriminating 
detectors. In this situation, according to Eq.~(\ref{f^q_iIdealDet}), 
 we have $P^{qrst}_{ijkl}=\delta_{qi}\delta_{rj}\delta_{sk}\delta_{tl}$   
and the quantum state (\ref{StateAfterImperfectMeasurement}) 
reduces to a single component, namely the pure 
state $\ket{\Phi_{1010}}$. According to (\ref{Result-For:Phi(ijkl)}) 
we obtain 
\begin{equation}\label{StateBeforePostselection}
\ket{\Phi_{1010}}=\frac{1}{\sqrt{2}}\left(\frac{\ket{1010}-\ket{0101}}{\sqrt{2}}
+\frac{\ket{0011}-\ket{1100}}{\sqrt{2}}\right)\;.
\end{equation}
Thus, apart from the expected Bell state $\ket{\psi^-}_{ad}\,$, we get 
another term superposed to it, namely a superposition of two photons 
being in the $a$ mode (in different polarizations of the latter) 
and no photons in the $d$ mode and vice versa. 
This is also understandable intuitively. There are three quantum 
alternatives contributing 
to the event $(qrst)=(1010)$ of the Bell measurement: 
({\em i}) each PDC source emits exactly one photon-pair; ($ii$) 
the \lq\lq first'' PDC source produces vacuum and the \lq\lq second'' 
source two (independent) photon pairs; ($iii$) the \lq\lq first'' PDC 
source produces two photon pairs and the \lq\lq second'' source vacuum. 
The probability of each of these three alternatives is proportional to
$\chi^4$, which also explains why the resultant normalized quantum state 
(\ref{StateBeforePostselection}) of the remaining modes $a$ and $d$ 
does not depend on $\chi$.
Only the alternative ($i$) leads to the desired Bell state
$\ket{\psi^-}_{ad}$, whereas the other two alternatives ($ii$) 
and ($iii$) entail the second term in (\ref{StateBeforePostselection}). 
For this feature see also \cite{Riedmatten2003Phys.Rev.A67}.

Hence, entanglement swapping performed with  PDC sources cannot 
herald Bell states in the outgoing modes with a 100\% 
probability, even if the detectors used for the Bell 
measurement are perfectly ideal. If, however, we are interested 
in {\em post-selected} correlations of detection events 
of the $a$ and $d$ modes, only the Bell state $\ket{\psi^-}_{ad}$ 
contributes to them. And as expected, the detection events will 
be perfectly anti-correlated.  
The second term in Eq.\ (\ref{StateBeforePostselection}) 
does not contribute to four-fold coincidences and thus can 
be ignored in terms of post-selection. Furthermore, to 
fulfill the commonly used relation (\ref{F-V-Relation}) 
between the fidelity and the experimentally observed 
visibility of four-fold coincidences, the former has to 
be calculated with the post-selected state. 
This means we have to project the quantum                                                  
state  $\hat{\rho}^{qrst}$ onto the subspace which corresponds to 
click events in both the $a$ and $d$ modes. Each of the 
outgoing modes has to have at least one photon. Introducing 
the projection operator   
\begin{equation}
\hat{\Pi}_{\mbox{\tiny
   postsel}}:=\left(\one_{a_{H},a_{V}}-\left(\ket{00}\bra{00}\right)_{a_{H},a_{V}}\right)\otimes
\left(\one_{d_{V},d_{H}}-\left(\ket{00}\bra{00}\right)_{d_{V},d_{H}}\right)\;,
\end{equation}
the post-selected quantum state is defined as: 
\begin{equation}
 \hat{\rho}^{qrst}_{\mbox{\tiny
    postsel}}:=\frac{
\hat{\Pi}_{\mbox{\tiny
    postsel}} 
\hat{\rho}^{qrst}
\hat{\Pi}_{\mbox{\tiny
    postsel}}}
{\Tr[\hat{\Pi}_{\mbox{\tiny
    postsel}} 
\hat{\rho}^{qrst}
\hat{\Pi}_{\mbox{\tiny
    postsel}}]}\;.
\end{equation}
The fidelity 
\begin{equation}
F_{\mbox{\tiny
    postsel}}:=\bra{\psi^-}\hat{\rho}^{qrst}_{\mbox{\tiny
    postsel}}\ket{\psi^-}\;.
\end{equation}
fulfills the relation (\ref{F-V-Relation}), 
(see~\cite{RiedmattenMarcikicHouwelingenTittelZbindenGisin2005}).

\begin{figure}[b] 
 \center{\includegraphics[width=0.65\linewidth]{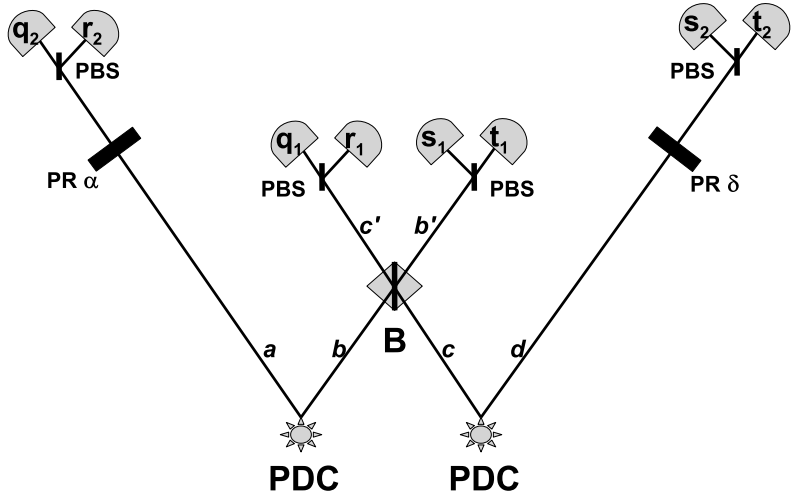}}
{\caption{ A scheme for entanglement verification in an entanglement 
swapping experiment. A Bell measurement with imperfect threshold detectors 
is performed on the $c$ and $b$ modes. A read-out $(q_1r_1s_1t_1)$ is 
recorded. Polarization rotators (PR) parameterized by 
angles $\alpha$ and $\delta$ are applied to the remaining 
$a$ and $d$ modes prior to polarization measurements using 
polarizing beamsplitters (PBS) and non-ideal threshold detectors. 
The readout of the second measurement 
is denoted by  $(q_2r_2s_2t_2)$. For a fixed polarization rotation of 
the $a$ mode,  $\alpha=\pi/4$,   
the observation of certain {\em four-fold} 
coincidences for varying rotation angles $0\le\delta\le \pi$ of the  $d$ mode 
reveals, via the visibility $V$, 
the strength of the polarization correlations between the $a$ and $d$ modes.
\label{Fig:Four-foldCoincidencesExperiment}}}
\end{figure}

Our scheme for entanglement verification via four-fold coincidence
measurements is very similar to the experimental setup   
in~\cite{PanBouwmeesterWeinfurterZeilinger1998}. It is illustrated in
Fig.~\ref{Fig:Four-foldCoincidencesExperiment}. We implement 
{\em variable} polarization measurements 
on both modes $a$ and $d$ by introducing {\em polarization rotators} (PR)
into their spatial paths prior to polarizing beam-splitters and 
threshold detectors. The polarizations of the  $a$ and $d$ modes 
are rotated by independently variable angles $\alpha$ and $\delta$. 
To avoid any confusion, let us agree upon the following meaning 
of the latter. The angles $\alpha$ and $\delta$ stand for rotation angles 
of polarization vectors in the real space and not of Bloch vectors 
on the Bloch sphere. Neither do they mean rotation angles of 
the $\lambda/2$-plates, which are used in an actual experiment to 
cause polarization rotations. The relation between these three 
different meanings is as follows. A rotation by angle $\alpha$ 
of a polarization vector in real space corresponds to a 
rotation by angle $\tilde{\alpha}=2\alpha$ of the Bloch 
vector on the  Bloch sphere and to a rotation by angle 
$\alpha_{\lambda/2}=\alpha/2$ of the $\lambda/2$-plate.

Polarization rotations prior to PBS separating horizontal 
and vertical polarizations, and photon-detections, are 
equivalent to polarization measurements in different 
bases in Hilbert space. The absolute angle of rotation in each of the modes determines the 
basis of the polarization measurement. The polarization correlations 
of the entangled state should, in the ideal-case scenario, depend 
only on the relative angle between the two polarization rotators.  
We choose to rotate the polarization of mode $a$ by a fixed angle $\alpha=\pi/4$ (i.e.\ $+45^\circ$)
and numerically calculate the probabilities for coincidences 
of detector clicks for measurements on the  $a$ and $d$ modes,  
for varying angles $\delta$ of polarization rotation of the 
$d$ mode, {\em given that} the Bell measurement has yielded the 
result $(1,0,1,0)$ {\em or} $(0,1,0,1)$. We denote the four 
detectors for the measurements on the  $a$ and $d$ modes by 
$D_a^+\,,D_a^-\,,D_d^-\,$ and $D_d^+$, corresponding 
to analyzing the $a$ mode along the $+45^\circ$-axis and 
 $-45^\circ$-axis and the $d$ mode along the 
variable polarization directions $-\delta$ and 
 $+\delta$, respectively. Furthermore, the four-tuple event 
$(q_2,r_2,s_2,t_2)$ of the measurements on the 
 $a$ and $d$ modes, cf.~Fig.~\ref{Fig:Four-foldCoincidencesExperiment}, 
corresponds to readouts of the four-tuple of detectors 
$(D_a^+,D_a^-,D_d^-,D_d^+)$.

In what follows, we provide the probability for four-fold coincidences 
using our model for practical entanglement swapping. 
Mathematically, the polarization rotators acting on the $a$ and $d$ modes 
are represented by the unitary operators
\begin{eqnarray}\label{Unitary1:PolarizationRotators}
\hat{U}_a(\tilde{\alpha})&=&\exp\left[i\tilde{\alpha}\hat{J}_a\right]\;,\\
\hat{U}_d(\tilde{\delta})&=&\exp\left[i\tilde{\delta}\hat{J}_d\right]\;,
\label{Unitary2:PolarizationRotators}
\end{eqnarray}
with $\tilde{\alpha},\tilde{\delta}\in\mathbb{R}$ and the generators of rotation 
\begin{eqnarray}\label{Generator1:PolarizationRotators}
\hat{J}_a&:=&\frac{1}{2}\left(\hat{a}_{\mbox{\tiny{V}}}^\dagger\hat{a}_{\mbox{\tiny{H}}}+\hat{a}_{\mbox{\tiny{V}}}\hat{a}_{\mbox{\tiny{H}}}^\dagger\right)\;,\\
\hat{J}_d&:=&\frac{1}{2}\left(\hat{d}_{\mbox{\tiny{V}}}^\dagger\hat{d}_{\mbox{\tiny{H}}}+\hat{d}_{\mbox{\tiny{V}}}\hat{d}_{\mbox{\tiny{H}}}^\dagger\right)\;.
\label{Generator2:PolarizationRotators}
\end{eqnarray}
Here the angles $\tilde{\alpha}$ and $\tilde{\delta}$ are rotation angles of the Bloch vectors 
on the Bloch sphere and not of the polarization vectors in real
space. As mentioned above, the former are related to the latter 
via $\tilde{\alpha}=2\alpha\;, \tilde{\delta}=2\delta$.

Given an event $(q_1r_1s_1t_1)$ of an imperfect Bell measurement on modes 
$b$ and $c$, 
the conditional probability that polarization measurements on the    
$a$ and $d$ modes would have yielded the result $(i_2j_2k_2l_2)$, 
if the detectors had been ideal, is given by 
\begin{eqnarray}
\mbox{prob}\left(i_2j_2k_2l_2|q_1r_1s_1t_1\right)
&=&\Tr\bigg[\left(\ket{i^{a_H}_2j^{a_V}_2k^{d_V}_2l^{d_H}_2}\bra{i^{a_H}_2j^{a_V}_2k^{d_V}_2l^{d_H}_2}\right)
\times\nonumber\phantom{\bigg]}\\ &&\times\phantom{\bigg[}
\left(\hat{U}_a(\tilde{\alpha})\otimes\hat{U}_d(\tilde{\delta}\,)
\hat{\rho}^{q_1r_1s_1t_1}
\Ud_a(\tilde{\alpha})
\otimes
\Ud_d(\tilde{\delta}\,)\right)
\bigg]\nonumber\\
&=&\sum_{i_1j_1k_1l_1}W^{i_1j_1k_1l_1}_{i_2j_2k_2l_2}(\tilde{\alpha},\tilde{\delta}\,)
P^{q_1r_1s_1t_1}_{i_1j_1k_1l_1}(\,\chi,\{\eta^{(1)}_{\nu}\},\{{\wp^{(1)}_{\mbox{\tiny{dc}}\nu}}\})
\end{eqnarray}
where we have used Eq.~(\ref{StateAfterImperfectMeasurement}) and 
introduced the transition probabilities
\begin{eqnarray}\label{TransitionProbability}
W^{i_1j_1k_1l_1}_{i_2j_2k_2l_2}(\tilde{\alpha},\tilde{\delta}\, )&:=&
\left|\bra{i^{a_H}_2j^{a_V}_2k^{d_V}_2l^{d_H}_2} \hat{U}_a(\tilde{\alpha})\otimes\hat{U}_d(\tilde{\delta}\,)
\ket{\Phi_{i_1j_1k_1l_1}}\right|^2\;.
\end{eqnarray}
The conditional probability to observe the event $(q_2r_2s_2t_2)$ with 
non-ideal, imperfect detectors, given an event $(q_1r_1s_1t_1)$ of a non-ideal
Bell measurement on modes $b$ and $c$, is denoted and calculated as:
 \begin{eqnarray}
&& Q^{q_1r_1s_1t_1}_{q_2r_2s_2t_2}\left(\chi,\{\eta^{(1)}_{\nu}\},\{\eta^{(2)}_{\nu}\},
\{{\wp^{(1)}_{\mbox{\tiny{dc}}\nu}}\},\{{\wp^{(2)}_{\mbox{\tiny{dc}}\nu}}\},
\tilde{\alpha},\tilde{\delta}\right):= 
\mbox{prob}\left(q_2r_2s_2t_2|q_1r_1s_1t_1\right)\nonumber\\
&&=\sum_{i_2j_2k_2l_2}\mbox{prob}(q_2r_2s_2t_2|i_2j_2k_2l_2)\,\mbox{prob}(i_2j_2k_2l_2|q_1r_1s_1t_1)\nonumber\\
&&=\sum_{i_2j_2k_2l_2}
p_{\eta_1^{(2)},\wp^{(2)}_{\mbox{\tiny{dc}}1}}(q_2|i_2)\:
p_{\eta_2^{(2)},\wp^{(2)}_{\mbox{\tiny{dc}}2}}(r_2|j_2)\:
p_{\eta_3^{(2)},\wp^{(2)}_{\mbox{\tiny{dc}}3}}(s_2|k_2)\:
p_{\eta_4^{(2)},\wp^{(2)}_{\mbox{\tiny{dc}}4}}(t_2|l_2)\:
\times\nonumber\\
&&\times\left(\sum_{i_1j_1k_1l_1}
W^{i_1j_1k_1l_1}_{i_2j_2k_2l_2}\big(\tilde{\alpha},\tilde{\delta}\, \big)
P^{q_1r_1s_1t_1}_{i_1j_1k_1l_1}\big(\,\chi,\{\eta^{(1)}_{\nu}\},\{{\wp^{(1)}_{\mbox{\tiny{dc}}\nu}}\}\big) \right)
\nonumber\\
&&= \sum_{i_2,j_2,k_2,l_2=0}^\infty \;\sum_{i_1,j_1,k_1,l_1=0}^\infty
p_{\eta_1^{(2)},\wp^{(2)}_{\mbox{\tiny{dc}}1}}(q_2|i_2)\:
p_{\eta_2^{(2)},\wp^{(2)}_{\mbox{\tiny{dc}}2}}(r_2|j_2)\:
p_{\eta_3^{(2)},\wp^{(2)}_{\mbox{\tiny{dc}}3}}(s_2|k_2)\:
p_{\eta_4^{(2)},\wp^{(2)}_{\mbox{\tiny{dc}}4}}(t_2|l_2)\:
\nonumber\\&&\hspace{3cm}\times 
W^{i_1j_1k_1l_1}_{i_2j_2k_2l_2}\big(\tilde{\alpha},\tilde{\delta}\,\big)P^{q_1r_1s_1t_1}_{i_1j_1k_1l_1}\big(\,\chi,\{\eta^{(1)}_{\nu}\},\{{\wp^{(1)}_{\mbox{\tiny{dc}}\nu}}\}\big)
\end{eqnarray}

For numerical simulations we need to know the transition probabilities 
(\ref{TransitionProbability}). An explicit analytic expression is 
provided in the appendix C. This result together with 
Eqs.~(\ref{AccordingToBayesTheoremGeneralized}) and 
(\ref{ResultFor:f^q_iForthresholddetectors1}-\ref{ResultFor:f^q_iForthresholddetectorsDef-h}) 
enable us to calculate numerically the probabilities for four-fold
coincidences. As explained above, we condition on obtaining the readout 
$(1_{c'_{H}},0_{c'_V},1_{b'_V},0_{b'_H})$
{\em or} 
$(0_{c'_{H}},1_{c'_V},0_{b'_V},1_{b'_H})$
in the Bell measurement. 
Given {\em either} of these two events occurs, regardless
which of them, we calculate the conditional probabilities of recording the
events $(1,0,1,0)$, $(0,1,1,0)$, $(0,1,0,1)$, and $(1,0,0,1)$,
using the four-tuple $(D_a^+,D_a^-,D_d^-,D_d^+)$ of imperfect 
threshold detectors in the polarization measurement on the $a$ and $d$ modes, 
depending on the varying angle of rotation $\delta$. 
 \begin{figure}[tb] 
 \center{\includegraphics[width=0.75\linewidth]{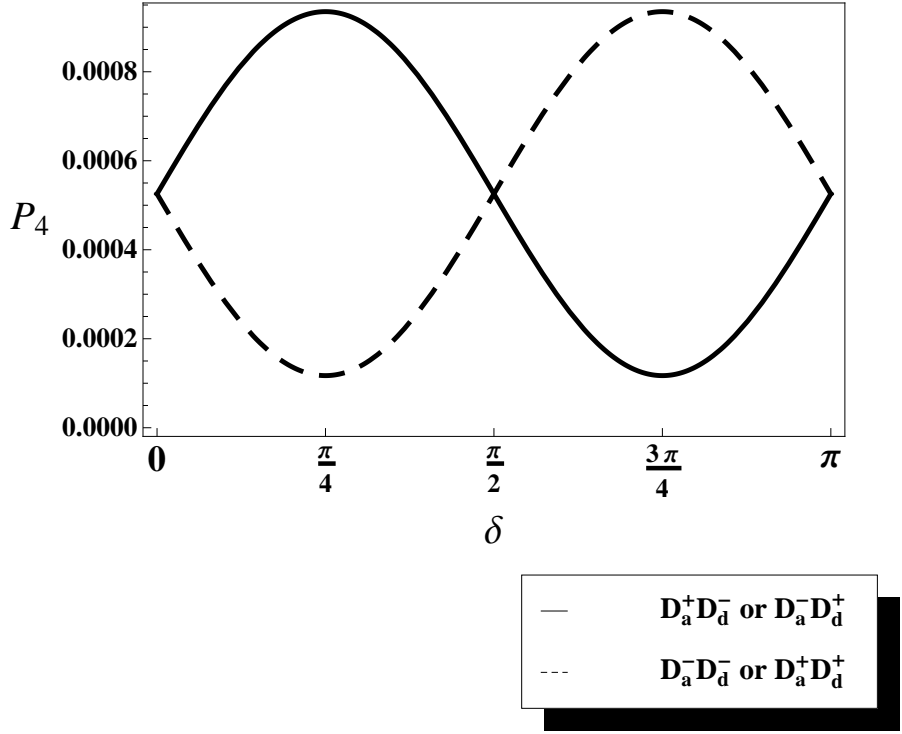}}
{\caption{ Simulated probability of {\em four-fold coincidence} $P_4$ 
depending on polarization rotation angle $\delta$, for a fixed 
rotation angle $\alpha=\pi/4$. Here $\alpha$ and $\delta$ represent 
angles of polarization rotations of the $a$ and $d$ modes in real space 
and not on the Bloch sphere. The experimental parameter values are 
given by $\chi\approx \sqrt{0,06}\approx 0.24$, 
$\eta^{(1)}_{1,2}=0.045$, $\eta^{(1)}_{3,4}=0.135$, 
$\eta^{(2)}_{1,2,3,4}=0.04$ and 
$\wp^{(1)}_{\mbox{\tiny{dc}}\,3,4}=1\times 10^{-5}, 
\wp^{(1)}_{\mbox{\tiny{dc}}\,1,2}=\wp^{(2)}_{\mbox{\tiny{dc}}\,1,2,3,4}=3\times 
10^{-5}$ in accordance with the conditions as found in the recent 
experiment~\cite{RiedmattenMarcikicHouwelingenTittelZbindenGisin2005}.  
The continuous sinusoidal curve refers 
to the probability  of the click coincidences \lq\lq $D^+_a$ {\em and }
$D^-_d$'' or  \lq\lq $D^-_a$ {\em and }
$D^+_d$''  , and the dashed  sinusoidal curve displays the probability of the
click coincidences  \lq\lq $D^-_a$ {\em and } $D^-_d$'' or 
\lq\lq $D^+_a$ {\em and } $D^+_d$'', respectively, {\em given that} the Bell measurement yielded 
the result \lq\lq $H$ {\em and } $V$'' {\em or} 
\lq\lq $V$ {\em and } $H$'' for the $c$ and $b$ mode, respectively. 
The  visibility is $V_{\mbox{\tiny th}}=77.7\%$.  
\label{Fig:Four-foldSinus}}}
\end{figure}

In order to compare our predictions with experimental entanglement swapping, 
we have to choose the same experimental parameter values $\chi$,
$\{\eta^{(1)}_{\nu}\}$, $\{\eta^{(2)}_{\nu}\}$,
$\{{\wp^{(1)}_{\mbox{\tiny{dc}}\nu}}\}$, $\{{\wp^{(2)}_{\mbox{\tiny{dc}}\nu}}\}$ 
as in the experiment. It is important to emphasize that the detector 
efficiencies $\eta$ of our theory are to be considered as {\em effective} 
efficiencies that also include transmission and other losses, 
cf.~Sec.~\ref{Sec:InefficientDetectors}.

\begin{figure}[bt]
\centering
 \center{\includegraphics[width=1.0\linewidth]{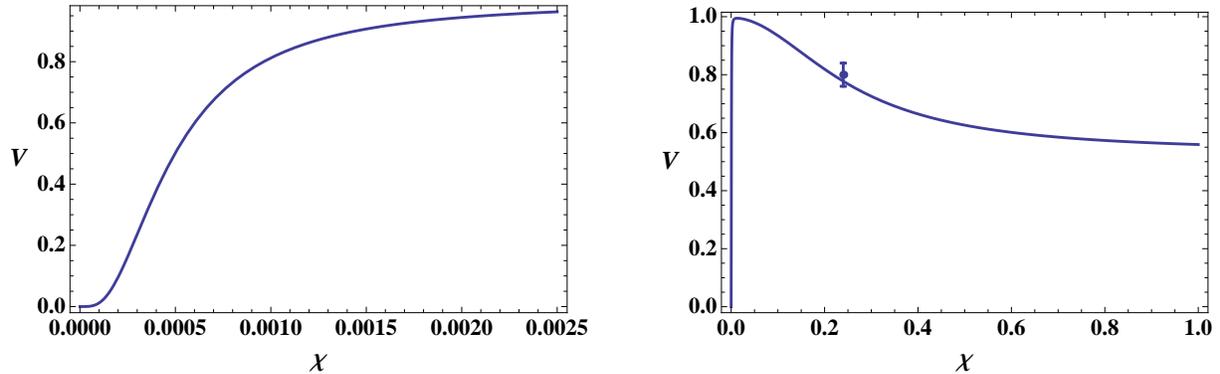}}
{\caption{
Visibility $V$ versus the square root of the photon-pair production rate    
$\chi$, for 
$\eta^{(1)}_{1,2}=0.045$, $\eta^{(1)}_{3,4}=0.135$,  
$\eta^{(2)}_{1,2,3,4}=0.04$ and 
$\wp^{(1)}_{\mbox{\tiny{dc}}\,3,4}=1\times 10^{-5},   
\wp^{(1)}_{\mbox{\tiny{dc}}\,1,2}=\wp^{(2)}_{\mbox{\tiny{dc}}\,1,2,3,4}=3\times 10^{-5}$. 
These values refer to the experimental situation of 
\cite{RiedmattenMarcikicHouwelingenTittelZbindenGisin2005}.  
The left image displays
the functional dependence for very small  $\chi$ 
values, the right image reveals the behavior 
for appreciably large  values of $\chi$.  
For $\chi\approx 0.24$, which 
is the value found in the 
experiment \cite{RiedmattenMarcikicHouwelingenTittelZbindenGisin2005},
our numerical simulation (curve) yields the visibility 
$V_{\mbox{\tiny th}}=77.7\%$.  The experimental result $V_{\mbox{\tiny
      exp}}=(80\pm 4)\%$ 
of  \cite{RiedmattenMarcikicHouwelingenTittelZbindenGisin2005} 
is depicted by means of an $\pm 4\%$ error bar. A very 
good agreement is achieved.  
Our result also clearly demonstrates that the chosen photon-pair production 
rate belongs to a region with a rapidly decreasing visibility, 
thus being already far beyond its optimal value. 
\label{Fig:Visibility}}}
\end{figure}

While we have developed our theory with entanglement swapping of polarization 
qubits as an intuitive example, it obviously also holds for any other
realization of qubits, 
e.g.\ time-bin qubits \cite{TittelWeihs2001}. 
In the following we compare the predictions of our theory with a recent
experiment on time-bin entanglement 
swapping~\cite{RiedmattenMarcikicHouwelingenTittelZbindenGisin2005}, 
which explicitly mentions all required parameters related to the PDC sources, 
transmission line and detectors. 
Fig.~\ref{Fig:Four-foldSinus} and Fig.~\ref{Fig:Visibility} demonstrate
results  of our numerical simulations of four-fold coincidences with 
reference to experimental conditions of  \cite{RiedmattenMarcikicHouwelingenTittelZbindenGisin2005}.  
The conditions of this experiment are given by the following approximate 
values: $\chi\approx \sqrt{0,06}\approx 0.24$, 
$\eta^{(1)}_{1,2}=0.045$, $\eta^{(1)}_{3,4}=0.135$, 
$\eta^{(2)}_{1,2,3,4}=0.04$ and 
$\wp^{(1)}_{\mbox{\tiny{dc}}\,3,4}=1\times 10^{-5}, 
\wp^{(1)}_{\mbox{\tiny{dc}}\,1,2}=\wp^{(2)}_{\mbox{\tiny{dc}}\,1,2,3,4}=3\times
10^{-5}$.
As expected, we get two complementary sine curves that are $90^{\circ}$ out 
of phase, one curve for anti-correlated polarization readouts 
(\lq\lq $D^+_a$ {\em and } $D^-_d$'' or \lq\lq $D^-_a$ {\em and } $D^+_d$'') 
and another  curve for correlated polarization readouts
(\lq\lq $D^-_a$ {\em and } $D^-_d$'' or \lq\lq $D^+_a$ {\em and } $D^+_d$''),
at the detectors for the  $a$ and $d$ modes. 
As anticipated, the probability to detect anti-correlated polarizations   
for the  $a$ and $d$ modes attains its maximum for $\delta=\alpha=\pi/4$ 
and its minimum for $\delta=\alpha+\pi/2=3\pi/4$. Complementary to
this,  the probability to detect correlated polarizations has its maximum 
for  $\delta=\alpha+\pi/2=3\pi/4$ and its minimum for 
$\delta=\alpha=\pi/4$. This is the characteristic entanglement property 
of the Bell state  $\ket{\Psi^-}_{ad}$. The calculated visibility amounts to  
 $V_{\mbox{\tiny th}}=77.7\%$. This is in respectably good agreement with the experimentally 
achieved  visibility $V_{\mbox{\tiny exp}}=(80\pm 4)\%$  
of  \cite{RiedmattenMarcikicHouwelingenTittelZbindenGisin2005}.

Fig.~\ref{Fig:Visibility} reveals the behavior of the visibility 
as a function of the square root of the photon-pair production rate $\chi$, 
for detector efficiencies and dark count probabilities chosen 
as given above. It is interesting 
to observe that the photon-pair production rate of 
the PDC sources used in this experiment, $\chi\approx 0.24$, 
lies far beyond its optimal value. 

We have also compared the predictions of our model with the experimental data from 
the polarization entanglement swapping experiment reported in
\cite{JenneweinWeihsPanZeilinger2002}. 
In this case, using $\chi$ =0.05 (which can be calculated from reported count
rates, 
transmission loss and detector quantum efficiencies), 
we find $V_{\mbox{\tiny th}}=98\%$. Note that the experimentally obtained visibility, 
$V_{\mbox{\tiny exp}}\approx 90\%$, is much smaller than the value obtained
from our model. 
This can easily be explained by taking into account that detector noise and
multi-pair emissions are not the only practical limitations that impact on 
the visibility. Other deficiencies include imperfect entanglement created 
by each source, even in the case where only one single pair is generated, 
and imperfect temporal overlap of modes $b$ and $c$ on the beamsplitter 
that realizes the Bell state measurement. We believe the latter to be the 
main limitation in this experiment. Hence, our model currently only provides 
an upper, yet useful bound on the visibility.

\begin{figure}[tb]
\centering
 \center{\includegraphics[width=1.0\linewidth]{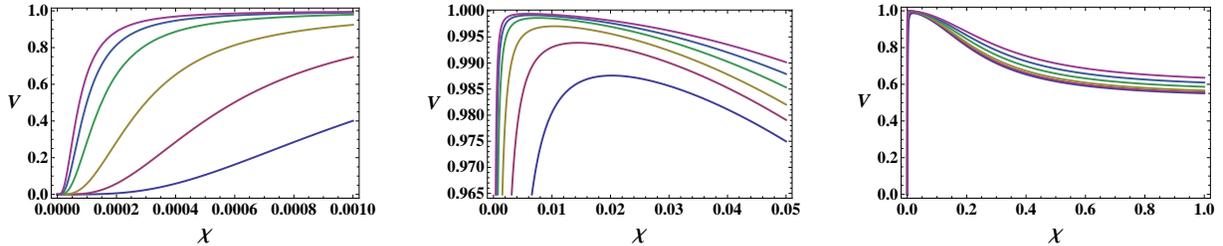}}
{\caption{ Visibility $V$ versus the square root of the photon-pair production
    rate $\chi$, for 
various detector efficiencies $\eta$ and the fixed dark count probability 
$\wp_{\mbox{\tiny{dc}}}=1\times 10^{-5}$. 
The function $V_{\eta,\wp_{\mbox{\tiny{dc}}}}(\chi)$ 
is plotted for six different efficiency values: $\eta=
0.025,\:0.05,\: 0.1,\: 0.2,\: 0.3,\: 0.4$, corresponding to the 
curves of lowest to highest visibility in all three 
diagrams. The left diagram displays the dependence 
of the function
$V_{\eta,\wp_{\mbox{\tiny{dc}}}}(\chi)$ on the parameter  $\eta$ 
for very small $\chi$ values, the middle 
diagram shows the neighborhood of the maxima 
 and the right diagram the behavior for 
high  $\chi$ values. The regions displayed in the first two 
plots are also included in the third diagram where 
they correspond to the steep increase of the visibility 
for very small $\chi$ values.
\label{Fig:VisibilityVersusChiForDifferentEtas}}}
\end{figure}
\begin{figure}[tb]
\centering
 \center{\includegraphics[width=1.0\linewidth]{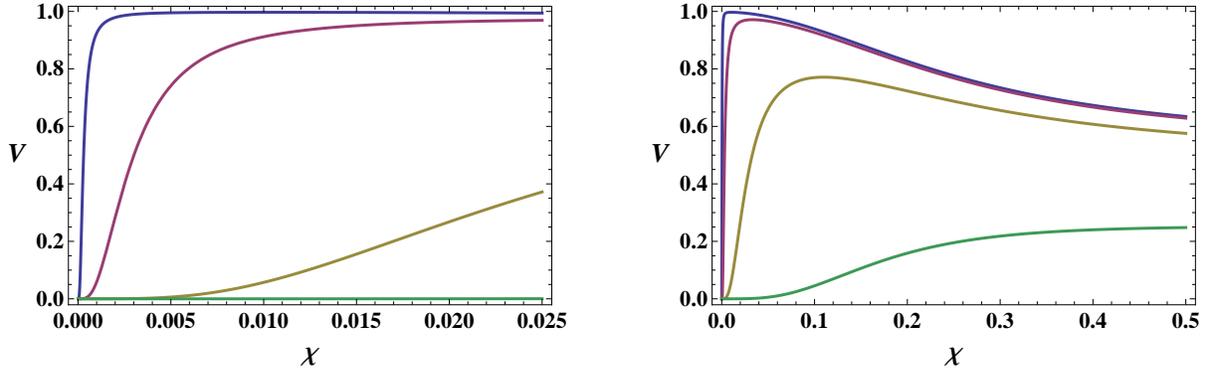}}
{\caption{ Visibility $V$ versus the square root of the photon-pair production
 rate $\chi$, for various detector dark count probabilities and the fixed 
efficiency $\eta=0.1$.  Plots of the function $V_{\eta,\wp_{\mbox{\tiny{dc}}}}(\chi)$ 
are shown for four different detector dark count probabilities: 
$\wp_{\mbox{\tiny{dc}}}=1\times 10^{-2},\:1\times10^{-3},\:1\times
10^{-4},\:1\times 10^{-5}$, 
corresponding to the curves of lowest to highest visibility in both 
diagrams. The left diagram displays the dependence 
of the function
$V_{\eta,\wp_{\mbox{\tiny{dc}}}}(\chi)$ on the parameter  $\wp_{\mbox{\tiny{dc}}}$ 
for small $\chi$ values, the right diagram the behavior for 
high  $\chi$ values.
\label{Fig:VisibilityVersusChiForDifferentDarkCounts}}}
\end{figure}
To analyze how entanglement is affected by detector imperfections, 
we have calculated the visibility $V_{\eta,\wp_{\mbox{\tiny{dc}}}}(\chi)$ 
as a function of $\chi$ 
for  various detector efficiencies $\eta$ and dark count 
probabilities  $\wp_{\mbox{\tiny{dc}}}$. To keep the analysis 
simple, we have assumed the same  efficiencies and dark count 
probabilities for all detectors involved in the experimental setting. 
The dependence on $\eta$ and  $\wp_{\mbox{\tiny{dc}}}$ is illustrated in
Fig.~\ref{Fig:VisibilityVersusChiForDifferentEtas} 
and  Fig.~\ref{Fig:VisibilityVersusChiForDifferentDarkCounts}. 
It is interesting to observe that, according to our theory, 
for reasonable low dark count probabilities ($\wp_{\mbox{\tiny{dc}}} \le
10^{-4}$) there is a region of $\chi$-values such that 
the visibility can be made very close to $100\%$ even 
for low detector efficiencies. Given any  
 detector efficiencies and dark counts, 
our model makes it possible to provide  
the optimal photon-pair production rate 
in order to achieve high entanglement 
after entanglement swapping. This is 
a valuable result. An important conclusion 
that can be drawn from our investigations is the 
feature that high  photon-pair production rates 
are counter-productive. One should not 
exceed values of about $\chi= 5\times 10^{-2}$. 
Higher $\chi$ values lead to a decreasing entanglement 
in the final quantum state after entanglement swapping 
and as such have an adverse impact. 
Yet, depending on the application, $V$ may not be the appropriate figure 
of merit to optimize.

First, in the here discussed proof-of-principle entanglement swapping 
experiments, the goal was to demonstrate a violation of the CHSH Bell 
inequality, rather than to maximize the violation. An important concern 
was thus to limit the time it took to take the data, generally many 
days. For the
experiment~\cite{RiedmattenMarcikicHouwelingenTittelZbindenGisin2005} 
we referred to in Fig.~\ref{Fig:Visibility}, 
this has resulted in source brightnesses 
$\chi$ exceeding the value required for maximum visibility.

As a second example where the visibility is not the relevant quantity to 
optimize, let us briefly consider QKD. Here, the figure of merit is the 
secret key rate, which scales in a non-trivial way with the visibility and 
with the number of detected coincidences \cite{Ma2007}. For the highest 
possible key rate, an optimal trade-off between the production rate of 
final entangled pairs, and visibility (i.e.\ amount of entanglement) has to 
be achieved. Hence, reducing the brightness of the PDC sources to a value 
that results in a maximum for $V$ does in general not lead 
to a maximum secret key rate.

\section{Conclusion}
\label{Conclusion}

Within the scope of our theory of practical entanglement swapping 
we have derived the actual quantum state of the remaining $a$ and $d$ 
modes depending on the result of a noisy Bell measurement with 
imperfect detectors and probabilistic sources. The main achievement consists in our 
ability to provide this quantum state for any given  
photon-pair production rate $\chi$, any given detector efficiency  
$\eta$ and any reasonable dark-count probability $\wp_{\mbox{\tiny{dc}}}$. 
In deriving our results we have made a distinction 
between photon-number discriminating detectors 
and threshold detectors. The calculated quantum state allows us to make predictions 
with regard to any quantity of interest. In particular, we can 
calculate the entanglement of the remaining modes, parameterized 
in terms of the visibility obtained in coincidence 
measurements, 
depending on the experimental parameters  $\chi$, $\eta$ and
$\wp_{\mbox{\tiny{dc}}}$. 

Predictions of our theory have proved to be in close accord with experimental 
entanglement swapping. Our numerical simulations of certain four-fold
coincidences between four threshold detectors 
demonstrate a remarkably good agreement with a recent entanglement swapping 
experiment reported in~\cite{RiedmattenMarcikicHouwelingenTittelZbindenGisin2005}.

Furthermore, our theory provides 
a very useful functional relation 
between entanglement, quantified, e.g., by the visibility 
of four-fold coincidences, and the square root of 
the photon-pair production rate $\chi$.
The latter is an essential  experimental quantity in 
long-distance quantum key distribution 
based on protocols that employ quantum relays and repeater.
The secret key rate, or equivalently the quantum bit error rate referred to as 
QBER, depend crucially on the photon-pair creation rate $\chi$. 
In our opinion, there sometimes is a tendency among 
scientists working on experimental QKD 
to aim at achieving  {\em brighter} PDC sources due to the 
prevailed conception that higher photon-pair production 
rates lead to higher sifted-key rates. 
Here it should be realized, however, that it is not the source brightness  
itself that matters, but rather the production rate of final entangled 
photon-pairs after the swapping protocol, with their quantum correlations 
being as close as possible to that of Bell states. 
Our results indicate that in some experiments 
the {\em optimal value} of $\chi$ might 
eventually be exceeded. The investigations presented in
section~\ref{ComparisonWithExperimentalSwapping} suggest 
that, for a visibility of at least $97\%$, the 
photon-pair creation rates should not be chosen much greater 
than $\chi=0.05$. Higher values decrease the 
entanglement present after swapping, due to undesired multi-pair 
events from the same source,  thus implying an increasing QBER. 
Hence, our results allow us to suggest 
the implications of the imperfections 
on schemes using entanglement swapping 
as a fundamental tool.

It should be emphasized, however, that, depending on the application,  
the amount of entanglement, as quantified by $V$,
is not always the appropriate quantity to be 
made maximal. In QKD, for instance, 
the figure of merit is given by the secret key rate $R_{\mbox{\tiny sec}}$, 
and the non-trivial dependence of the latter on $\chi$ and the visibility  
$V$, cf.\ e.g.~\cite{Ma2007}, suggests that the optimal $\chi$ values with respect to QKD are not 
necessarily those which yield the highest visibility of entanglement swapping.  
Nevertheless, we believe that the methods of the present work will prove very 
useful in finding the optimal photon-pair production 
rates with regard to achieving optimal secret-key rates in long distance,
quantum repeater based QKD. 
As a first step into this direction we will have to generalize 
our results to several concatenated noisy entanglement swappings,   
and then examine the scaling properties with the number of these segments in the iteration.

In our future considerations we also plan to investigate the issue 
whether the optimal $\chi$ can be shifted to higher values by using 
photon-number discriminating detectors instead of threshold detectors 
for the Bell-state measurement. To understand the intuition, let us 
briefly discuss the events where a coincidence detection in two 
threshold detectors in modes $b'$ and $c'$  is interpreted as a projection 
onto a Bell state. For threshold detectors, a fraction of these cases 
originates from two photons impinging on one detector, and one photon 
impinging on the other detector. These undesired but not identifiable 
events result in a reduction of the swapped entanglement, i.e.\  
visibility. Unit efficiency photon-number discriminating detectors would 
allow identifying and discarding these events, which leads to a higher 
visibility for equal source brightness, or, conversely, allows 
increasing the brightness while keeping the visibility constant. 
And so would do imperfect photon-number discriminating detectors 
to some extent. This conjecture is worth analyzing in view of promising technological 
advancements in research on photon-counting detectors 
\cite{Rosenberg2005A,Rosenberg2005B}. Obviously, though, 
if the PDC sources are too bright, events where only one click is 
detected at each output will be rare, and the production rate 
of final entangled pairs after the entanglement swapping operation 
will be low.

\vspace*{3mm}
\noindent{\bf Acknowledgment:} 
The authors acknowledge the support of NSERC, iCORE, CIFAR, MITACS, and General
Dynamics Canada in preparing this work. We also thank Peter Marzlin and 
Thomas Jennewein for helpful discussions.

\section{Appendix}
\subsection{Dark count probability
  $\boldsymbol{\wp_{\mbox{\tiny{dc}}}}$}

Throughout the paper we characterize dark counts in terms of 
the experimental parameter 
$\wp_{\mbox{\tiny{dc}}}$, which is defined to be the probability
of a dark count event per detection window in a non-ideal threshold detector. 
Within our detector model, Eq.~(\ref{ProbabilityForDarkCount}) 
provides $\wp_{\mbox{\tiny{dc}}}$ as a function of 
the parameter $\,r\,,$ or equivalently the pseudo temperature 
of the thermal source (\ref{ThermalSource}). To derive this relation, 
let us consider our detector model as depicted in Fig.~\ref{DetectorModel}   
and at first assume the ideal photon-detector $D_{\mbox{\scriptsize id}}$ to 
be photon-number discriminating. Furthermore, we also suppose the 
signal mode to be in the vacuum state: $\hat{\rho}_{\mbox{\tiny sig}}=
\vac\bra{\mbox{vac}}$. Then, the probability 
of detecting $k$ dark count photons in the detector $D_{\mbox{\scriptsize
    id}}$ is obtained by applying the beam-splitter transformation
$\hat{U}_{\mbox{\tiny BS}}(\eta)$ to the state 
$\vac\bra{\mbox{vac}}\otimes\hat{\rho}_{\mbox{\tiny T}}$, followed by 
tracing over the reflected mode and projecting the transmitted 
mode onto the Fock state $\ket{k}$, and finally tracing over the latter mode.   
Hereby \lq\lq reflection'' and \lq\lq transmission'' refer to the signal mode.
Thus, with the notation 
\begin{equation}
\label{DefinitionOf-b(.)}
\tilde{b}(\eta,r):=\eta\tanh^2 r\;,
\end{equation}
we have:
 \begin{eqnarray}
\mbox{prob}(\mbox{\it \lq\lq k dark count photons"})&=& 
\Tr_{\mbox{\tiny trans}}\left[\hat{\Pi}_k\Tr_{\mbox{\tiny refl}}\left[
\hat{U}_{\mbox{\tiny BS}}(\eta)
\left(\vac\bra{\mbox{vac}}\otimes\hat{\rho}_{\mbox{\tiny T}}\right)
\hat{U}^\dagger_{\mbox{\tiny BS}}(\eta)
\right]\hat{\Pi}_k\right]\;\nonumber\\
&=& \frac{1}{\cosh^2 r}\sum_{n=k}^\infty {n\choose k}
\left[\tanh^2 r\right]^n\eta^{n-k}(1-\eta)^k\;\nonumber\\
&=&\frac{1}{\cosh^2 r}\left(\frac{1-\eta}{\eta}\right)^k\sum_{n=k}^\infty {n\choose k}
\left[\tilde{b}(\eta,r)\right]^n\nonumber\\
&=&\frac{1}{\cosh^2 r}\left(\frac{1-\eta}{\eta}\right)^k\left[\tilde{b}(\eta,r)\right]^k\sum_{\nu=0}^\infty {\nu+k\choose k}
\left[\tilde{b}(\eta,r)\right]^\nu\;,
\end{eqnarray}
where we used $\hat{\Pi}_k=\ket{k}\bra{k}$ and the 
definition (\ref{ThermalSource}) of the thermal state $\hat{\rho}_{\mbox{\tiny
    T}}$. Since $0\le \tilde{b}(\eta,r)<1$, we may 
apply the Lemma 
\begin{equation}
\sum_{\nu=0}^\infty {\nu+k\choose k}
q^\nu=\frac{1}{(1-q)^{k+1}}\quad\mbox{for}\quad 0\le q<1\;,
\end{equation}
which can be found in \cite{Stanley1999}.
Hence, 
\begin{equation}
\mbox{prob}(\mbox{\it \lq\lq k dark count photons"})=
\frac{1}{\cosh^2 r}\left(\frac{1-\eta}{\eta}\right)^k
\left[\frac{\tilde{b}(\eta,r)}{1-\tilde{b}(\eta,r)}\right]^k
\frac{1}{1-\tilde{b}(\eta,r)}\;.
\end{equation}
To obtain the probability
for a dark count event in a photon-detector that 
does not discriminate between photon numbers, 
we have to sum over all integers $k\not=0$: 
 \begin{eqnarray}
\wp_{\mbox{\tiny{dc}}}&=&\sum_{k=1}^\infty 
\mbox{prob}(\mbox{\it \lq\lq k dark count photons"})\nonumber\\
&=&\frac{1}{[1-\tilde{b}(\eta,r)]\cosh^2 r}
\sum_{k=1}^\infty
\left[\frac{(1-\eta)\tilde{b}(\eta,r)}{\eta(1-\tilde{b}(\eta,r))}\right]^k\;.
\end{eqnarray} 
Since 
\begin{equation}
\theta:=\frac{(1-\eta)\tilde{b}(\eta,r)}{\eta[1-\tilde{b}(\eta,r)]}=
\frac{(1-\eta)\tanh^2 r}{1-\eta\tanh^2 r}<1\;,
\end{equation}
we finally get 
 \begin{eqnarray}
\wp_{\mbox{\tiny{dc}}}&=&
\frac{1}{\left[1-\tilde{b}(\eta,r)\right]\cosh^2 r}\left(\sum_{k=0}^\infty
 \theta^k-1\right)\nonumber\\
&=&\frac{1}{\left[1-\tilde{b}(\eta,r)\right]\cosh^2 r}\left(\frac{1}{1-\theta}-1\right)\nonumber\\
&=&\frac{(1-\eta)\tanh^2 r}{1-\eta\tanh^2 r}\;,
\label{ProbabilityForDarkCountInAppendix}
\end{eqnarray}
which is the result (\ref{ProbabilityForDarkCount}). Using this relation, 
we can express the function (\ref{DefinitionOf-b(.)}) in terms of 
$\eta$ and $\wp_{\mbox{\tiny{dc}}}$ rather than $\eta$ and $r$:
 \begin{eqnarray}
\label{Eq:btilde-->b}
b(\eta,\wp_{\mbox{\tiny{dc}}})
&:=&\tilde{b}\left(\eta,r(\eta,\wp_{\mbox{\tiny{dc}}})\right)\nonumber\\
&=&\eta\tanh^2r=\left[ 1+\frac{1-\eta}{\eta
    \wp_{\mbox{\tiny{dc}}}}\right]^{-1}\;,
\end{eqnarray}
which is Eq.~(\ref{Eq:b-in-terms-of-pdc}).

\subsection{Conditional probabilities
  $\boldsymbol{p_{\eta,\wp_{\mbox{\tiny{dc}}}}(q|i)}$}

The conditional probability $p_{\eta,\wp_{\mbox{\tiny{dc}}}}(q|i)$ to detect $q$ photons given 
that $i$ photons are incident upon a photon-number discriminating detector 
with efficiency $\eta$ and dark count probability $\wp_{\mbox{\tiny{dc}}}$ 
constitutes a major ingredient in our derivation of the quantum state after 
a realistic entanglement swapping operation. Here we derive its closed-form 
expression (\ref{ResultForProb(q|i)}) using our detector model 
as depicted in Fig.~\ref{DetectorModel}. Assuming 
 the signal mode to be in the Fock state
$\hat{\rho}_{\mbox{\tiny sig}}=\ket{i}\bra{i}$, 
according to Eq.~(\ref{ResultForProb(q|rho)withDarkCounts})  
the conditional probability of interest is given as: 
\begin{equation}\label{Eq1:DerConditionalProbability-q-i}
p_{\eta,\wp_{\mbox{\tiny{dc}}}}(q|i)
=\Tr_{\mbox{\tiny
trans}}\left[\hat{\Pi}_q\Tr_{\mbox{\tiny refl}}\left[
\hat{U}_{\mbox{\tiny BS}}(\eta)
\left(\ket{i}\bra{i}
\otimes\hat{\rho}_{\mbox{\tiny T}}\right)
\hat{U}^\dagger_{\mbox{\tiny BS}}(\eta)
\right]\hat{\Pi}_q\right]\;.
\end{equation}
In what follows, we expand on the calculation of this expression. 
Using the relation (\ref{ProbabilityForDarkCount}), 
we expressed in (\ref{Eq1:DerConditionalProbability-q-i}) the dependence 
on the conditions of the thermal source 
in terms of the label $\wp_{\mbox{\tiny{dc}}}$ rather than 
the pseudo temperature of the source  
(or equivalently its parameter $r$).
We will, however, initially use the 
notation $p_{\eta,r}(q|i)$ in the 
calculation below, and later 
show how the dark count probability 
$\wp_{\mbox{\tiny{dc}}}$ emerges 
in the final result. 

Let $\hat{c},\;\hat{c}^{\dagger}$ 
and $\hat{e},\;\hat{e}^{\dagger}$ denote the annihilation and 
creation operators of the signal mode and the thermal mode 
(fictitious environment), respectively. Initially, 
their common quantum state, denoted by $\hat{\varrho}^{\mbox{\tiny in}}_{\mbox{\tiny sig,T}}$, 
is a product state, namely 
$\hat{\varrho}^{\mbox{\tiny in}}_{\mbox{\tiny sig,T}}=\ket{i}\bra{i}
\otimes\hat{\rho}_{\mbox{\tiny T}}$, where 
$\hat{\rho}_{\mbox{\tiny T}}$ is given by 
(\ref{ThermalSource}). We can express this input state 
as an operator functional of creation and annihilation operators:
\begin{equation}
\hat{\varrho}^{\mbox{\tiny in}}_{\mbox{\tiny sig,T}}=
\hat{\varrho}^{\mbox{\tiny in}}_{\mbox{\tiny
    sig,T}}\left[\hat{c},\hat{c}^{\dagger},\hat{e},\hat{e}^{\dagger}
 \right]=
\frac{1}{\cosh^2r}\sum_{n=0}^\infty\frac{\tanh^{2n}r}{i!n!}
(\hat{c}^{\dagger})^i(\hat{e}^{\dagger})^n
\ket{\mbox{vac}}\bra{\mbox{vac}}\hat{e}^n\hat{c}^i\;.
\end{equation}
Combining the signal with the thermal mode at a beamsplitter with 
transmittance $\eta$ leads to an entangled quantum state. 
In the Schr{\"o}dinger picture, in which the photonic operators 
$\hat{c},\hat{c}^{\dagger},\hat{e},\hat{e}^{\dagger}$ remain unchanged, 
the beam-splitter transformation 
is represented by the unitary operator $\hat{U}_{\mbox{\tiny BS}}(\eta)$ 
which acts on input quantum states of the  beamsplitter. Its action 
corresponds to replacing the creation and annihilation operators 
in the functional $\hat{\varrho}^{\mbox{\tiny in}}_{\mbox{\tiny
    sig,T}}\left[\hat{c},\hat{c}^{\dagger},\hat{e},\hat{e}^{\dagger}
 \right]$ by new operators according to the rule~\cite{VogelWelsch2006}:
 \begin{equation}
{\hat{c}\choose \hat{e}}\rightarrow B_{\eta}^\dagger{\hat{c}\choose
  \hat{e}}\quad,\quad
{\hat{c}^\dagger\choose \hat{e}^\dagger}\rightarrow B_{\eta}^T{\hat{c}^\dagger\choose
  \hat{e}^\dagger}\;,
\end{equation}
where 
\begin{equation}
B_{\eta}=
\begin{pmatrix} 
 \sqrt{\eta}     & \sqrt{1-\eta} \\ 
  - \sqrt{1-\eta} & \sqrt{\eta}
\end{pmatrix}
\end{equation}
is the beam-splitter transformation matrix consisting of the transmission
and reflection coefficients $T=\sqrt{\eta}$ and $R=\sqrt{1-\eta}$.
Applying the beam-splitter transformation yields the following entangled
quantum state: 
\begin{eqnarray}\label{Eq:StateAfterBeamsplitterInTheDetectormodel}
\hat{\varrho}^{\mbox{\tiny in}}_{\mbox{\tiny sig,T}}  \stackrel{\mbox{\scriptsize BS}}\longrightarrow 
\hat{\varrho}^{\mbox{\tiny out}}_{\mbox{\tiny sig,T}}
&:=&
\hat{U}_{\mbox{\tiny BS}}(\eta)\hat{\varrho}^{\mbox{\tiny in}}_{\mbox{\tiny sig,T}}
\hat{U}^\dagger_{\mbox{\tiny BS}}(\eta)\nonumber\\
&=&\hat{U}_{\mbox{\tiny BS}}(\eta)
\left(\ket{i}\bra{i}\otimes\hat{\rho}_{\mbox{\tiny T}}\right)
\hat{U}^\dagger_{\mbox{\tiny BS}}(\eta)\nonumber\\
&=&\frac{1}{\cosh^2r}\sum_{n=0}^\infty\frac{\tanh^{2n}r}{i!\,n!}
\left(\sqrt{\eta}\,\hat{c}^{\dagger}-
\sqrt{1-\eta}\,\hat{e}^{\dagger}\right)^i
\left(\sqrt{1-\eta}\,\hat{c}^{\dagger}+
\sqrt{\eta}\,\hat{e}^{\dagger}\right)^n\nonumber\\
&&\times\,\ket{\mbox{vac}}\bra{\mbox{vac}}
\left(\sqrt{1-\eta}\,\hat{c}+
\sqrt{\eta}\,\hat{e}\right)^n
\left(\sqrt{\eta}\,\hat{c}-
\sqrt{1-\eta}\,\hat{e}\right)^i 
\nonumber\\
&=&\frac{1}{\cosh^2r}\sum_{n=0}^\infty\frac{\tanh^{2n}r}{i!\,n!}
\sum_{\nu=0}^i\sum_{\nu'=0}^i\sum_{\mu=0}^n\sum_{\mu'=0}^n
{ i \choose  \nu}
{ i \choose  \nu'}
{ n \choose  \mu}
{ n \choose  \mu'} \nonumber\\
&&\times(-1)^{\nu+\nu'}  \left(\sqrt{\eta}\right)^{2n+\nu+\nu'-\mu-\mu'}
 \left(\sqrt{1-\eta}\right)^{2i-\nu-\nu'+\mu+\mu'}\nonumber\\
&&\times 
\left(\hat{c}^{\dagger}\right)^{\nu+\mu}
\left(\hat{e}^{\dagger}\right)^{i+n-\nu-\mu}
\ket{\mbox{vac}}\bra{\mbox{vac}}
\hat{e}^{i+n-\nu'-\mu'}
\hat{c}^{\nu'+\mu'} \;. 
\end{eqnarray}
According to Eq.~(\ref{Eq1:DerConditionalProbability-q-i}) the 
conditional probability $p_{\eta,r}(q|i)$ is obtained by 
ignoring the reflected mode and projecting 
the transmitted mode onto the Fock state $\ket{q}$, 
and finally taking the trace of the resultant 
non-normalized state. This can be rewritten as:
\begin{equation}\label{Eq2:DerConditionalProbability-q-i}
p_{\eta,r}(q|i) =\Tr\left[\hat{\varrho}^{\mbox{\tiny out}}_{\mbox{\tiny
      sig,T}}
\left(\hat{\Pi}_q^{\mbox{\tiny
trans}}\otimes\one^{\mbox{\tiny
refl}}   \right)\right]\;.
\end{equation}
By inserting (\ref{Eq:StateAfterBeamsplitterInTheDetectormodel}) into  
(\ref{Eq2:DerConditionalProbability-q-i}) and realizing that in the 
Schr{\"o}dinger picture, which we use here, the mode $c$ in the 
equation (\ref{Eq:StateAfterBeamsplitterInTheDetectormodel}) above  
corresponds to the transmitted and the mode $e$ to 
the reflected mode, respectively,
we get: 
\begin{eqnarray}
p_{\eta,r}(q|i) 
&=&\frac{1}{\cosh^2r}\sum_{n=0}^\infty\frac{\tanh^{2n}r}{i!\,n!}
\sum_{\nu=0}^i\sum_{\nu'=0}^i\sum_{\mu=0}^n\sum_{\mu'=0}^n
{ i \choose  \nu}
{ i \choose  \nu'}
{ n \choose  \mu}
{ n \choose  \mu'} \nonumber\\
&&\times\,(-1)^{\nu+\nu'} \left(\sqrt{\eta}\right)^{2n+\nu+\nu'-\mu-\mu'}
 \left(\sqrt{1-\eta}\right)^{2i-\nu-\nu'+\mu+\mu'}\nonumber\\
&&\times\, \sqrt{(\nu+\mu)!}\sqrt{(\nu'+\mu')!} 
\sqrt{(i+n-\nu-\mu)!}\sqrt{(i+n-\nu'-\mu')!} \nonumber\\
&&\times \,\delta_{q,\,\nu+\mu}\delta_{q,\;\nu'+\mu'}
 \delta_{i+n-\nu-\mu\,,\,i+n-\nu'-\mu'}\;\nonumber\\
&=&\frac{1}{\cosh^2r}\sum_{n=0}^\infty\frac{\tanh^{2n}r}{i!\,n!}
\sum_{\mu=0}^n\sum_{\mu'=0}^n
{ i \choose  q-\mu}
{ i \choose  q-\mu'}
{ n \choose  \mu}
{ n \choose  \mu'}
(-1)^{\mu+\mu'}\nonumber\\
&&\times\,\eta^{n+q}(1-\eta)^{i-q}\left(\frac{1-\eta}{\eta}\right)^{\mu+\mu'}
q!(i+n-q)!\nonumber\\
&=&\frac{1}{\cosh^2r}\left(\frac{\eta}{1-\eta}\right)^q(1-\eta)^i
\sum_{n=0}^\infty\left[\eta\tanh^{2}r \right]^n
\frac{q!(i+n-q)!}{i!\,n!}\left[ \Omega(\eta,i,q,n)\right]^2\;.
\label{Eq:ConditProbabilityWithOmega}
\end{eqnarray}
In the last step we have introduced the definition
\begin{equation}\label{Eq:OmegaFunction}
\Omega(\eta,i,q,n):=\sum_{\mu=0}^n{ i \choose  q-\mu}{ n \choose  \mu}
\left(\frac{\eta-1}{\eta}\right)^{\mu}\;.
\end{equation}
It is important to recognize that $ \Omega(\eta,i,q,n)=0$ unless 
$i\ge q-n$. This is intuitively understandable: the number $q$ 
of detected photons in $D_{\mbox{\scriptsize id}}$ cannot be 
greater than the sum $i+n$ of incident signal and thermal 
photons. To proceed, we have to distinguish two cases: 
$i\ge q$ and $i<q$. It is easily seen that in the case $i\ge q$ we have 
\begin{equation}\label{Eq:Omega1}
\Omega(\eta,i,q,n)={ i \choose
  q}\, _2F_1\left(-n,-q\,;i-q+1\,;\frac{\eta-1}{\eta}\right)\;,
\end{equation}
where  $_2F_1(\cdot,\cdot;\cdot;\cdot)$ is the Hypergeometric function
as defined in (\ref{Eq:DefHypergeometricFunction}). In the second case, 
$q>i$, we obtain:
\begin{eqnarray}
\Omega(\eta,i,q,n)&=&
\sum_{\mu=q-i}^n{ i \choose  q-\mu}{ n \choose  \mu}
\left(\frac{\eta-1}{\eta}\right)^{\mu} \nonumber\\
&=&\sum_{\kappa=0}^{n-q+i}{ i \choose  i-\kappa}{ n \choose \kappa +q-i}
\left(\frac{\eta-1}{\eta}\right)^{\kappa+q-i} \nonumber\\
&=&
\left(\frac{\eta-1}{\eta}\right)^{q-i}
\;\sum_{\kappa=0}^{n-q+i}{ i \choose  \kappa}{ n \choose n-q+i-\kappa}
\left(\frac{\eta-1}{\eta}\right)^{\kappa}
 \nonumber\\
&=&
\left(\frac{\eta-1}{\eta}\right)^{q-i}
{ n \choose n-q+i}
\sum_{\mu=0}^{n-q+i}{ i \choose  \mu}
{ n \choose n-q+i}^{-1}
{ n \choose n-q+i-\mu}
\left(\frac{\eta-1}{\eta}\right)^{\mu}
 \nonumber\\
&=&
\left(\frac{\eta-1}{\eta}\right)^{q-i}
{ n \choose q-i}\;_2F_1\left(-i,q-i-n;q-i+1;\frac{\eta-1}{\eta}\right)\;.
\label{Eq:Omega2}
\end{eqnarray}
We insert (\ref{Eq:Omega1}) into (\ref{Eq:ConditProbabilityWithOmega}) and 
utilize the notation (\ref{DefinitionOf-b(.)}) to arrive at the following 
interim result for $i\ge q$:
\begin{eqnarray}
p_{\eta,r}(q|i) 
&=&
 \frac{1}{\cosh^2r}\left(\frac{\eta}{1-\eta}\right)^q(1-\eta)^i 
\sum_{n=0}^\infty {i \choose q}{i-q +n \choose i-q}
\left[\tilde{b}(\eta,r)\right]^n\nonumber\\&&\times\, \left[\, _2F_1\left(-n,-q\,;i - q +1\,;\frac{\eta-1}{\eta}\right)\right]^2\;.
\label{Eq:ConditProbabilityInterimResult:i>q}
\end{eqnarray}
For the $q>i$ case, we insert (\ref{Eq:Omega2}) into
(\ref{Eq:ConditProbabilityWithOmega}) and obtain: 
\begin{eqnarray}
p_{\eta,r}(q|i) 
&=&
 \frac{1}{\cosh^2r}
\left(\frac{1-\eta}{\eta}\right)^{q-i}\eta^i 
\sum_{n=q-i}^\infty {q \choose i}{n \choose q-i}
\left[\tilde{b}(\eta,r)\right]^n\nonumber\\&&\times\, \left[\, _2F_1\left(-i,q-i-n;q-i+1;\frac{\eta-1}{\eta}\right)\right]^2\;\nonumber\\
&=&
 \frac{1}{\cosh^2r}
\left(\frac{1-\eta}{\eta}\right)^{q-i}\eta^i\, \left[\tilde{b}(\eta,r)\right]^{q-i}
\sum_{m=0}^\infty {q \choose i}{q-i+m \choose q-i}
\left[\tilde{b}(\eta,r)\right]^m\nonumber\\&&\times\, \left[\, _2F_1\left(-i,-m\,;q-i+1\,;\frac{\eta-1}{\eta}\right)\right]^2\;\nonumber\\
&=&
 \frac{1}{\cosh^2r}
\left[\frac{1-\eta}{\eta}\,\tilde{b}(\eta,r) \right]^{q-i}\eta^i
\sum_{n=0}^\infty {q \choose i}{q-i+n \choose q-i}
\left[\tilde{b}(\eta,r)\right]^n\nonumber\\&&\times\, \left[\, _2F_1\left(-n,-i\,;q-i+1\,;\frac{\eta-1}{\eta}\right)\right]^2\;.
\label{Eq:ConditProbabilityInterimResult:q>i}
\end{eqnarray}
In the last step, after renaming the index of summation, $m\rightarrow n$, we
made use of the permutation-symmetry property of Hypergeometric functions: 
\begin{equation}
_2F_1(\alpha,\beta;\gamma;z)=\,_2F_1(\beta,\alpha;\gamma;z)\;.
\end{equation}
In this way a partial symmetry between the two results 
(\ref{Eq:ConditProbabilityInterimResult:i>q}) and 
(\ref{Eq:ConditProbabilityInterimResult:q>i}) 
is achieved. Please note that the result 
(\ref{Eq:ConditProbabilityInterimResult:q>i})  
is valid also for $q=i$, as in this case the right-hand side
expressions of (\ref{Eq:ConditProbabilityInterimResult:i>q}) and 
(\ref{Eq:ConditProbabilityInterimResult:q>i}) coincide.

We would now like to 
replace the dependence on the parameter $r$ of the thermal 
source by a dependence on the dark count probability. The 
latter emerges in the following manner. We replace the function 
$\tilde{b}(\eta,r)$ by $b(\eta,\wp_{\mbox{\tiny{dc}}})$ 
according to Eq.~(\ref{Eq:btilde-->b}) and use Eq.~(\ref{ProbabilityForDarkCount}) 
to make the replacement
\begin{equation}
\frac{1}{\cosh^2r}=\frac{(1-\eta)(1-\wp_{\mbox{\tiny{dc}}})}{1-\eta(1-\wp_{\mbox{\tiny{dc}}})}\;.
\end{equation}
Finally, by introducing the definition (\ref{Def:G-Function}), 
we arrive at the result provided in Eq.~(\ref{ResultForProb(q|i)}).

\subsection{Transition probabilities 
 $\boldsymbol{W^{i_1j_1k_1l_1}_{i_2j_2k_2l_2}(\tilde{\alpha},\tilde{\delta} )}$}

In this appendix we provide an explicit analytic expression for 
the transition probabilities (\ref{TransitionProbability}). Their 
definition involves unitary transformations representing polarization rotators 
for the $a$ and $d$ modes.  
According to Eqs.~(\ref{Unitary1:PolarizationRotators})-(\ref{Generator2:PolarizationRotators})
the corresponding unitary operators are given by: 
\begin{eqnarray}\label{Unitary1alt:PolarizationRotators}
\hat{U}_a(\tilde{\alpha})&=&\exp\left[i\frac{\tilde{\alpha}}{2}\left(\hat{a}_{\mbox{\tiny{V}}}^\dagger\hat{a}_{\mbox{\tiny{H}}}+\hat{a}_{\mbox{\tiny{V}}}\hat{a}_{\mbox{\tiny{H}}}^\dagger\right)\right]\;,\\
\hat{U}_d(\tilde{\delta})&=&\exp\left[i\frac{\tilde{\delta}}{2}\left(\hat{d}_{\mbox{\tiny{V}}}^\dagger\hat{d}_{\mbox{\tiny{H}}}+\hat{d}_{\mbox{\tiny{V}}}\hat{d}_{\mbox{\tiny{H}}}^\dagger\right)\right]\;.
\label{Unitary2alt:PolarizationRotators}
\end{eqnarray}
To find an explicit formula for
$W^{i_1j_1k_1l_1}_{i_2j_2k_2l_2}(\tilde{\alpha},\tilde{\delta} )$ it is very useful 
to \lq\lq disentangle'' these exponentials using a generalized 
Baker-Cambell-Hausdorf formula. Let us define
\begin{equation}
\hat{S}_+:=\hat{a}_{\mbox{\tiny{V}}}^\dagger\hat{a}_{\mbox{\tiny{H}}}\;,\quad
\hat{S}_-:=\hat{a}_{\mbox{\tiny{V}}}\hat{a}_{\mbox{\tiny{H}}}^\dagger\;,\quad
\hat{S}_z:=\frac{1}{2}\left(\hat{a}_{\mbox{\tiny{V}}}^\dagger\hat{a}_{\mbox{\tiny{V}}}-
  \hat{a}_{\mbox{\tiny{H}}}^\dagger\hat{a}_{\mbox{\tiny{H}}}  \right)\;.
\end{equation}
These operators form a bosonic representation of the su(2) Lie algebra, 
as they obey the commutation relations of generators of the latter. 
According to \cite{Ban1993,PuriBook}, the following 
decomposition formula holds for exponential functions 
of the generators of the su(2) Lie algebra: 
\begin{equation}
\exp\left[i\tilde{\alpha}\left(\theta_+\hat{S}_++\theta_z\hat{S}_z+\theta_-\hat{S}_-\right)\right]
=\exp\left(f_+(\tilde{\alpha})\hat{S}_+\right)\exp\left(f_z(\tilde{\alpha})\hat{S}_z\right)\exp\left(f_-(\tilde{\alpha})\hat{S}_-\right)
\end{equation}
where the functions $f_{\pm}(\tilde{\alpha}),f_z(\tilde{\alpha})$ are given by:
\begin{eqnarray}
f_\pm(\tilde{\alpha})&:=&\frac{i\theta_\pm}{\Gamma_1}\frac{
\sin(\Gamma_1\tilde{\alpha})}{\cos(\Gamma_1\tilde{\alpha})-i\theta_z\sin(\Gamma_1\tilde{\alpha})/(2\Gamma_1)}\;,\\
f_z(\tilde{\alpha})&:=&-2\ln\left[\cos(\Gamma_1\tilde{\alpha})-i\frac{\theta_z}{2\Gamma_1} \sin(\Gamma_1\tilde{\alpha})\right]\;,
\end{eqnarray}
with 
\begin{equation}
\Gamma_1^2:=\theta_+\theta_-+\frac{\theta_z^2}{4}\;.
\end{equation}
We apply this decomposition formula to
Eq.~(\ref{Unitary1alt:PolarizationRotators}). In our case we have
$\theta_+=\theta_-=1/2$ and $\theta_z=0$, yielding $\Gamma_1=1/2\;$ and 
\begin{eqnarray}
f_\pm(\tilde{\alpha})&=&i\tan\frac{\tilde{\alpha}}{2}\;,\\
f_z(\tilde{\alpha})&=&-2\ln\left(\cos\frac{\tilde{\alpha}}{2}\right)\;.
\end{eqnarray}
Thus, Eq.~(\ref{Unitary1alt:PolarizationRotators}) becomes:
\begin{eqnarray}\label{Unitary1alt2:PolarizationRotators}
\hat{U}_a(\tilde{\alpha})&=&\exp\left[i\tan\left(\frac{\tilde{\alpha}}{2}\right) 
\hat{a}_{\mbox{\tiny{V}}}^\dagger\hat{a}_{\mbox{\tiny{H}}}\right]
\exp\left[-\ln\left(\cos\frac{\tilde{\alpha}}{2}\right)
\hat{a}_{\mbox{\tiny{V}}}^\dagger\hat{a}_{\mbox{\tiny{V}}}\right]\nonumber\\&&
\times\exp\left[\ln\left(\cos\frac{\tilde{\alpha}}{2}\right)
\hat{a}_{\mbox{\tiny{H}}}^\dagger\hat{a}_{\mbox{\tiny{H}}}\right]
\exp\left[i\tan\left(\frac{\tilde{\alpha}}{2}\right) \hat{a}_{\mbox{\tiny{V}}}\hat{a}_{\mbox{\tiny{H}}}^\dagger\right]\;.
\end{eqnarray}
In a similar way we obtain a decomposition of $\hat{U}_d(\tilde{\delta})$ in
Eq.~(\ref{Unitary2alt:PolarizationRotators}). The result reads: 
\begin{eqnarray}\label{Unitary2alt2:PolarizationRotators}
\hat{U}_d(\tilde{\delta})&=&\exp\left[i\tan\left(\frac{\tilde{\delta}}{2}\right) 
\hat{d}_{\mbox{\tiny{V}}}^\dagger\hat{d}_{\mbox{\tiny{H}}}\right]
\exp\left[-\ln\left(\cos\frac{\tilde{\delta}}{2}\right)
\hat{d}_{\mbox{\tiny{V}}}^\dagger\hat{d}_{\mbox{\tiny{V}}}\right]\nonumber\\&&
\times\exp\left[\ln\left(\cos\frac{\tilde{\delta}}{2}\right)
\hat{d}_{\mbox{\tiny{H}}}^\dagger\hat{d}_{\mbox{\tiny{H}}}\right]
\exp\left[i\tan\left(\frac{\tilde{\delta}}{2}\right) \hat{d}_{\mbox{\tiny{V}}}\hat{d}_{\mbox{\tiny{H}}}^\dagger\right]\;.
\end{eqnarray}
We use these expressions in (\ref{TransitionProbability}) to derive the
transition probabilities. A straightforward but lengthy  calculation yields 
the following result:
\begin{equation}
W^{i_1j_1k_1l_1}_{i_2j_2k_2l_2}(\tilde{\alpha},\tilde{\delta}\, )
=\left|A^{i_1j_1k_1l_1}_{i_2j_2k_2l_2}(\tilde{\alpha},\tilde{\delta} \,)
\right|^2
\end{equation}
with
\begin{eqnarray}  
A^{i_1j_1k_1l_1}_{i_2j_2k_2l_2}(\tilde{\alpha},\tilde{\delta}\,)
&:=& \bra{i^{a_H}_2j^{a_V}_2k^{d_V}_2l^{d_H}_2} \hat{U}_a(\tilde{\alpha})\otimes\hat{U}_d(\tilde{\delta})
\ket{\Phi_{i_1j_1k_1l_1}}\nonumber\\
&=&
\frac{1}{(\sqrt{2})^{i_1+j_1+k_1+l_1}
\sqrt{i_1!j_1!k_1!l_1!}}\sum_{\mu=0}^{i_1}\sum_{\nu=0}^{j_1}
\sum_{\kappa=0}^{k_1}\sum_{\lambda=0}^{l_1}
(-1)^{\mu+\nu}
{i_1 \choose \mu}
{j_1 \choose \nu}
{k_1 \choose \kappa}
{l_1 \choose \lambda}
\nonumber\\
&&\times\delta_{\mu+\nu+\kappa+\lambda\,,\,i_2+j_2}\,\delta_{i_1+j_1+k_1+l_1\,,\,i_2+j_2+k_2+l_2}
\nonumber\\
&&\times
\sqrt{(\mu+\lambda)!(\nu+\kappa)!(i_1+l_1-\mu-\lambda)!(j_1+k_1-\nu-\kappa)!}
\nonumber\\
&&\times
\sum_{n_a=0}^{\mbox{\scriptsize
    min}\{j_2,\nu+\kappa\}}\;\sum_{n_d=0}^{\mbox{\scriptsize
    min}\{k_2,j_1+k_1-\nu-\kappa\}}
\left[\cos\frac{\tilde{\alpha}}{2}\right]^{i_2+j_2-2n_a}\left[\cos\frac{\tilde{\delta}}{2}\right]^{k_2+l_2-2n_d}\nonumber\\
&&\times
\left\{\frac{\left[i\tan\frac{\tilde{\alpha}}{2}\right]^{j_2+\nu+\kappa-2n_a}
\left[i\tan\frac{\tilde{\delta}}{2}\right]^{k_2+j_1+k_1-\nu-\kappa-2n_d}}
{(j_2-n_a)!(\nu+\kappa-n_a)!(k_2-n_d)!(j_1+k_1-\nu-\kappa-n_d)!}\right\}
\nonumber\\
&&\times
\left[\prod_{m_1=1}^{j_2-n_a} (n_a+m_1)\right]^{\frac{1}{2}}
\left[\prod_{m_2=1}^{\nu+\kappa-n_a} (n_a+m_2)\right]^{\frac{1}{2}}
\nonumber\\
&&\times
\left[\prod_{m_3=1}^{j_2-n_a} (i_2+m_3)\right]^{\frac{1}{2}}
\left[\prod_{m_4=1}^{\nu+\kappa-n_a }(i_2+j_2 - \nu -\kappa +m_4)\right]^{\frac{1}{2}}
\nonumber\\
&&\times
\left[\prod_{m_5=1}^{k_2-n_d}(n_d+m_5)\right]^{\frac{1}{2}}
\left[\prod_{m_6=1}^{j_1+k_1-\nu-\kappa-n_d}(n_d+m_6)\right]^{\frac{1}{2}}
\nonumber\\
&&\times
\left[\prod_{m_7=1}^{k_2-n_d}(l_2+m_7)\right]^{\frac{1}{2}}
\left[\prod_{m_8=1}^{j_1+k_1-\nu-\kappa-n_d}  (k_2 + l_2 - j_1 - k_1+\nu+\kappa+m_8)\right]^{\frac{1}{2}}\nonumber\\
\end{eqnarray}

\end{document}